\documentclass[10pt,french,aps,showkeys,showpacs,nofootinbib]{revtex4-1}
\usepackage[T1]{fontenc}
\usepackage[utf8]{inputenc}
\usepackage{textcomp}
\usepackage{amsmath}
\usepackage{amssymb}
\usepackage{stackrel}
\RequirePackage{color}

\makeatletter
\usepackage[utf8]{inputenc}
\usepackage{todonotes}
\usepackage[bottom]{footmisc}
\usepackage{graphicx}
\usepackage{dcolumn}
\usepackage{bm}
\usepackage{todonotes}
\usepackage{hyperref}

\usepackage[scale=0.75]{geometry}
\makeatother

\usepackage{babel}
\makeatletter
\addto\extrasfrench{
   \providecommand{\fg}{\ifdim\lastskip>\z@\unskip\fi~\frqq}
}

\makeatother
\begin{document}

\title{ Applications of the Klein-Gordon equation in the Feshbach-Villars representation in the non-inertial cosmic string space-time}

\author{Abdelmalek Bouzenada}
\email{abdelmalekbouzenada@gmail.com; abdelmalek.bouzenada@univ-tebessa.dz}
\affiliation{Laboratory of theoretical and applied Physics,~\\
Echahid Cheikh Larbi Tebessi University, Algeria}
 
\author{Abdelmalek Boumali}
\email{boumali.abdelmalek@gmail.com}
\affiliation{Laboratory of theoretical and applied Physics,~\\
Echahid Cheikh Larbi Tebessi University, Algeria}
 
\author{Edilberto O. Silva}
\email{edilberto.silva@ufma.br}
\affiliation{Departamento de Física,~\\
 Universidade Federal do Maranhão, 65085-580 São Luís, Maranhão, Brazil}

\date{\today}
\begin{abstract}
We study the relativistic quantum motion of a spineless particle using the Feshbach-Villars (FV) formalism in the spinning cosmic string spacetime. The movement equations are derived using the first-order FV formulation of the Klein-Gordon (KG) equation. We apply the equation of motion (a) to study the motion of the particle confined to a rigid-wall potential, (b) motion in the presence of a Coulomb-type potential, and (c) particle interacting with the Feshbach-Villars oscillator (FVO). The energy levels and wave functions are obtained for the three cases. Our study focused on the impact of rotation and curvature on the energy levels of the particle.
\end{abstract}


\maketitle

\section{Introduction }

The study of how the gravitational field affects the dynamics of quantum mechanical systems holds significant importance. On the one hand, the theory of general relativity (GR), as proposed by Einstein \Citep{key-1}, provides a compelling explanation of gravity as a geometric property of space-time. It establishes that the classical gravitational field manifests space-time curvature and has successfully predicted phenomena such as gravitational waves \Citep{key-2} and black holes \Citep{key-3}. On the other hand, quantum mechanics (QM) serves as the framework for comprehending the behavior of particles at the microscopic level \Citep{key-4}. Quantum mechanics, often manifested through quantum field theory, has proven highly successful in elucidating the interactions among small particles and the emergence of three fundamental forces: weak, strong, and electromagnetic interactions \Citep{key-5}. However, the endeavor to construct a unified theory that reconciles general relativity with quantum mechanics, known as quantum gravity, has encountered numerous obstacles and unresolved technical challenges \Citep{key-6,key-7}.

A key approach to developing a theory that bridges the gap between gravity and relativistic quantum mechanics is by generalizing certain aspects of particle dynamics in flat Minkowski space to curved geometries \Citep{key-8,key-9}. This allows for the extension of this formulation, providing a comprehensive understanding of how the gravitational field influences the quantum behavior of relativistic particles. By incorporating arbitrarily curved backdrops, a broader framework can be constructed, shedding light on the intricate interplay between gravity and quantum mechanics at the relativistic level.

The versatility of this technique allows for its adaptation to address a wide array of models involving the concept of curvature, enabling the integration of a broader set of predictions concerning the values of macroscopic observables. These predictions play a crucial role in facilitating meaningful experimental verification of various phenomenological consequences, particularly in the domains of astrophysics and cosmology. Additionally, gaining a comprehensive understanding of the thermodynamic behavior of relativistic particles, with due consideration of gravitational effects \Citep{key-10,key-11}, presents a promising avenue for exploration. Moreover, studying related characteristics, such as fundamental statistical variables, further enhances the potential to draw important and necessary conclusions within the context of unraveling the enigmatic quantum behavior of gravity.

Topological defects (domain walls, cosmic strings, monopoles, and
textures) have been a focus of intense research over the last few
decades, and they continue to be one of the most active disciplines
in condensed matter physics, cosmology, astrophysics, and elementary
particle models. These structures are thought to have originated due to the Kibble mechanism \citep{key-12,key-13,key-14}, in which defects occur in symmetry-breaking phase transitions during the early universe's cooling \Citep{key-15,key-16}. The particular flaw in question is cosmic strings ( for more reviews, we refer the
reader to \Citep*{key-17}). These items (whether static or revolving) can have discernible impacts. These, for example, offer a viable method for seeding galaxy formation and the gravitational lensing effects. Furthermore, by investigating cosmic strings and their characteristics, we may learn a lot about particle physics at very high energies in many settings. Moreover, the potential that cosmic strings may act like superconducting wires has been proposed in current physics, with fascinating implications.

The harmonic oscillator (HO) has long been recognized as an essential instrument in many fields of theoretical physics \Citep{key-18}. It is a well-studied, absolutely solvable model that may be used to examine numerous complicated issues within the context of quantum mechanics \Citep{key-19}. addition relativistic extension of the quantum harmonic oscillator gives a useful model for understanding a wide range of molecular, atomic, and nuclear interactions. Moreover, the property of having a complete set of exact analytical solutions when working with such a model can give birth to radically different interpretations of many mathematical and physical events, and hence related applications can be realized via the underlying formulation.

The behavior of various relativistic quantum systems is now acknowledged to be critically dependent on the so-called Dirac oscillator (DO) as It\^{o} et al. \citep{key-20} specified in previous developments of spin-1/2 particle dynamics with a linear trajectory. They demonstrated that the non-relativistic limit of this system leads to an ordinary harmonic oscillator with a substantial spin-orbit coupling term. Actually, Moshinsky and Szczepaniak \citep{key-21} demonstrated that the DO might be derived from the free Dirac equation by introducing an external linear potential through a minimum replacement of the momentum
operator $\hat{p}_{\mu}\rightarrow \hat{p}_{\mu}-iM\omega\beta x_{\mu}$, with $\mu=0,1,2,3, \ldots$. It is worth noting that, in addition to the theoretical focus on researching the DO, useful insights may be achieved by examining physical interpretation, which is undoubtedly necessary for comprehending many relevant applications \cite{Silva1,Silva2,Silva3,Silva4,Silva5,boumali1,boumali2}.

Inspired by the DO, a similar formalism for the case of bosonic particles was presented, and it was dubbed a Klein-Gordon oscillator (KGO) \citep{key-23,key-24}. Several researchers have lately been working on the covariant version of this model in curved spacetimes and in various configurations. There are various contributions to the issue of relativistic quantum movements of scalar and vector particles under gravitational influences caused by different curved space-time geometries. For example, the
problem of the interaction between KGO coupled harmonically with topological defects in Kaluza-Klein theory is examined in Ref. \Citep{key-24}. Ref. \Citep{key-25} investigates the relativistic quantum dynamics of spin-0 particles in a spinning cosmic string space-time with Coulomb-type scalar and vector potentials. Additionally, rotating effects on the scalar field in cosmic string space-time, space-time with space-like dislocation, and space-time with spiral dislocation have been examined in Ref. \Citep{key-26}. The authors of Ref. \Citep{key-27} recently examined the KGO in cosmic string space-time and investigated the effects of the rotating frame and non-commutativity in momentum space.
Additionally, Ref. \Citep{key-28} investigated the Relativistic scalar charged particle in a rotating cosmic string space-time with Cornell-type potential and Aharonov-Bohm effect. Efforts have been made by several writers \Citep{key-30,key-31,key-32,boumali3,boumali4,boumali5}
to explore the relativistic spin-0, spin-1 bosons, and spin-1/2 fermions
wave functions and their time evolution using the Hamiltonian form,
i.e., possessing Schrodinger-type equations. In this regard, the so-called FV equations \Citep{key-33,key-34} are of special
relevance. These equations were originally designed to allow for a
relativistic single-particle interpretation of the second-order KG
equation. In the latter scenario, FV equations result from dividing the KG wave function into two components to produce an equation with
a first-order time derivative. Several investigations have been made 
in recent decades with the goal of examining the relativistic dynamical
features of single particles and solving their wave equations using
the FV technique (e.g., Refs. \Citep{key-35,key-36,key-37,key-38,key-39,key-40,key-41,key-42,key-43,key-44}
and other related references cited therein).

As mentioned in Ref. \citep{key-44}, the technique for introducing
a scalar potential into the KG equation is as follows. In
In the same way, the electromagnetic 4-vector potential is introduced.
This is achieved by modifying the momentum operator$p_{\alpha}=i\partial_{\alpha}$ to $p_{i}\rightarrow p_{i}-qA_{i}\left(x\right)$. Another method was described in Ref. \citep{key-45}, which included a modification in the mass of the particle as
$M\rightarrow M+S(\mathbf{r},t)$, where $S(\mathbf{r},t)$
indicates the scalar potential. In recent decades, researchers have
investigated the behavior of a Dirac particle in the presence of a
static scalar potential and a Coulomb potential \citep{key-45} as
well as a relativistic scalar particle in cosmic string spacetime
\citep{key-46}. Bouzenada et al. \citep{key-48} investigate the FVO case in spinning cosmic string space-time and discuss
some findings (thermal properties and density of this system). This same model has also been studied in the cosmic string space-time with a screw dislocation in the presence of coulomb-type potential \citep{key-51}. 

The structure of this paper is as follows. In Sec. \ref{sec2}, we present a brief review of the VF formalism emphasizing the relevant properties that are used in our work. In. Sec. \ref{sec3}, in the spinning cosmic string spacetime. In Sec. \ref{sec4}, we apply the KG equation in the VF representation to three physical models: (a) we study the case where the particle is confined to a hard-wall potential, (b) motion in the presence of a Coulomb-type potential, and (c) particle interacting with the KG oscillator. Our conclusions are presented in Sec. \ref{sec5}. Throughout the paper, we use natural units $\hbar=c=G=K_{B}=1$, and our metric
convention is $(+,-,-,-$).

\section{An overview of the Feshbach-Villars representation}\label{sec2}

This section focuses on the relativistic quantum depiction of a spin-0 particle moving through Minkowski space-time with the metric tensor
$\eta_{\mu\nu}=\text{diag}\left(1,-1,-1,-1\right)$. We begin with the conventional covariant KG equation to a scalar massive particle given by \Citep{key-43,key-44}
\begin{equation}
\left(\eta^{\mu\nu}D_{\mu}D_{\nu}+M^{2}\right)\Psi(x,t)=0,\label{eq:1}
\end{equation}
where $M$ is the mass of the particle, and the minimally-coupled covariant derivative is represented as $D_{\mu}=i\left(p_{\mu}-eA_{\mu}\right)$. Here, $p_{\mu}=\left(E,-p_{i}\right)$ denotes the canonical four-momentum, $A_{\mu}=\left(A_{0},-A_{i}\right)$ represents the electromagnetic four-potential, and $e$ represents the electric charge.

It is important to note that at this juncture, it can be emphasized that Equation \eqref{eq:1} has the capability to be restated in the form of a Hamiltonian equation, involving the first derivative with respect to time, akin to a Schrödinger-type equation
\begin{equation}
\mathcal{H}\Phi\left(x,t\right)=i\,\frac{\partial}{\partial t}\Phi(x,t).\label{eq:2}
\end{equation}
The Hamiltonian $\mathcal{H}$ can be achieved using the FV linearization technique, which involves transforming Eq. (\ref{eq:1}) into a first-order differential equation with respect to time. To achieve this, we introduce the two-component wave function \Citep{key-43,key-44}
\begin{equation}
\Phi(x,t)=\left(\begin{array}{c}
\phi_{1}(x,t)\\
\phi_{2}(x,t)
\end{array}\right)=\frac{1}{\sqrt{2}}\left(\begin{array}{c}
1+\frac{i}{M}\mathcal{D}\\
1-\frac{i}{M}\mathcal{D}
\end{array}\right)\psi(x,t),\label{eq:3}
\end{equation}
where $\psi(x,t)$ satisfies the KG wave equation, and $\mathcal{D}=\frac{\partial}{\partial t}+ieA_{0}(x)$. 
The transformation given in Eq. \eqref{eq:3} involves introducing wave functions that fulfill the specified conditions
\begin{equation}
\psi=\phi_{1}+\phi_{2},\qquad i\mathcal{D}\psi=M\left(\phi_{1}-\phi_{2}\right).
\end{equation}
To facilitate our subsequent examination, it is advantageous to express $\phi_{1}$ and $\phi_{2}$ as
\begin{align}
\phi_{1} & =\frac{1}{2M}\left[M+i\frac{\partial}{\partial t}-eA_{0}\right]\psi, \\
\phi_{2} & =\frac{1}{2M}\left[M-i\frac{\partial}{\partial t}+eA_{0}\right]\psi.
\end{align}
Substituting the equations of both $\phi_{1}$ and $\phi_{2}$ in \eqref{eq:3} and \eqref{eq:1} , we get
\begin{align}
\left[i\frac{\partial}{\partial t}-eA_{0}\right]\left(\phi_{1}+\phi_{2}\right) & =M\left(\phi_{1}-\phi_{2}\right),\\
\left[i\frac{\partial}{\partial t}-eA_{0}\right]\left(\phi_{1}-\phi_{2}\right) & =\left[\frac{\left(p_{i}-eA_{i}\right)^{2}}{M}+M\right]\left(\phi_{1}+\phi_{2}\right).
\end{align}
By adding and subtracting these two equations, we arrive at a system of interconnected first-order differential equations with respect to time
\begin{align}
\frac{\left(p_{i}-eA_{i}\right)^{2}}{2M}\left(\phi_{1}+\phi_{2}\right)+\left(M+eA_{0}\right)\phi_{1} & =i\frac{\partial\phi_{1}}{\partial t},\label{eq:8a}\\
-\frac{\left(p_{i}-eA_{i}\right)^{2}}{2M}\left(\phi_{1}+\phi_{2}\right)-\left(M-eA_{0}\right)\phi_{2} & =i\frac{\partial\phi_{2}}{\partial t}.\label{eq:8}
\end{align}
By employing Eqs. \eqref{eq:8a} and  \eqref{eq:8}, the Hamiltonian for a scalar particle under the influence of electromagnetic interaction in the FV formalism can be expressed as
\begin{equation}
\mathcal{H}_{FV}=\left(\tau_{z}+i\tau_{y}\right)\frac{\left(p_{i}-eA_{i}\right)^{2}}{2M}+M\tau_{z}+eA_{0}(x).\label{eq:9}
\end{equation}
The Pauli matrices in the conventional $2\times2$ form, denoted by $\tau_{i},\left(i=x,y,z\right)$, are defined as
\begin{equation}
\tau_{x}=\left(\begin{array}{cc}
0 & 1\\
1 & 0
\end{array}\right),\quad\tau_{y}=\left(\begin{array}{cc}
0 & -i\\
i & 0
\end{array}\right),\quad\tau_{z}=\left(\begin{array}{cc}
1 & 0\\
0 & -1
\end{array}\right).
\end{equation}
It is important to highlight that the Hamiltonian \eqref{eq:9} satisfies the condition of generalized Hermiticity. In other words, the Hamiltonian $\mathcal{H}$ is considered pseudo-Hermitian if there exists an invertible, Hermitian, linear operator $\beta$ such that the following condition is met: $\mathcal{H}^{\dagger}=\beta\mathcal{H}\beta^{-1}$ \citep{key-44}. Thus, we write 
\begin{equation}
\mathcal{H}_{FV}=\tau_{z}\mathcal{H}_{FV}^{\dagger}\tau_{z},\qquad\mathcal{H}_{FV}^{\dagger}=\tau_{z}\mathcal{H}_{FV}\tau_{z}.
\end{equation}
In the case of free particle propagation, where no interaction is considered (i.e., $A_{\mu}=0$), the one-dimensional FV Hamiltonian can be simplified to the following form:
\begin{equation}
\mathcal{H}_{0}=\left(\tau_{z}+i\tau_{y}\right)\frac{p_{x}^{2}}{2M}+M\tau_{z}. \label{eq:12}
\end{equation}
The solutions to the free Hamiltonian, which do not depend on time, correspond to stationary states. Assuming a solution in the form mentioned in Ref. \Citep{key-35}, we write
\begin{equation}
\Phi\left(x,t\right)=\Phi\left(x\right)e^{-iEt}=\left(\begin{array}{c}
\phi_{1}\left(x\right)\\
\phi_{2}\left(x\right)
\end{array}\right)e^{-iEt},\label{eq:13}
\end{equation}
with $E$ being the energy of the particle. Thus, Eq. \eqref{eq:2} can then be written as
\begin{equation}
\mathcal{H}_{0}\Phi\left(x\right)=E\Phi\left(x\right). \label{eq:14}
\end{equation}
The one-dimensional FV equation for a free relativistic spin-0 particle is an alternative Schrödinger-type equation to the KG equation. Its purpose is to provide a different perspective on the dynamics of the particle. In the subsequent analysis, this approach will be employed to derive solutions for wave equations in curved space-time, specifically focusing on the Non-inertial cosmic string. By utilizing this framework, a deeper understanding of the behavior of particles in the presence of curved space-time can be achieved.

\section{Quantum Dynamics of Spin-0 Particle in Non-inertial Cosmic String Space-time: The FV Representation}\label{sec3}

The equations governing the behavior of a scalar particle in a Riemannian spacetime, defined by the metric tensor $g_{\mu\nu}$, can be obtained by reformulating the Klein-Gordon equation. This approach, extensively discussed in various textbooks \citep{key-9,key-10,key-40,key-41,key-48,key-49,key-52}, allows for a deeper understanding of the dynamics of particles in curved spacetime. The KG equation is given by
\begin{equation}
\left(\frac{1}{\sqrt{-g}}\partial_{\mu}\left(\sqrt{-g}g^{\mu\nu}\partial_{\nu}\right)+M^{2}-\xi R\right)\Phi(x,t)=0.\label{eq:15}
\end{equation}
Here, the Laplace-Beltrami operator $\square=\frac{1}{\sqrt{-g}}\partial_{\mu}\left(\sqrt{-g}g^{\mu\nu}\partial_{\nu}\right)$ plays a crucial role. It represents the differential operator used in the KG equation in the curved space-time of the cosmic string, whose line element is defined later. Additionally, $\xi$ denotes a real dimensionless coupling constant, and $R$ corresponds to the Ricci scalar curvature defined by $R=g^{\mu\nu}R_{\mu\nu}$, with $R_{\mu\nu}$ representing the Ricci curvature tensor. The inverse metric tensor is denoted by $g^{\mu\nu}$, and $g=\det\left(g_{\mu\nu}\right)$ is the determinant of the metric tensor. These quantities collectively contribute to formulating the KG equation in the cosmic string space-time.

In the pursuit of understanding the quantum dynamics of spin-0 particles in space-time influenced by the non-inertial effects of a cosmic string, our current objective is to determine the equation of motion governing their motion. To accomplish this, we will employ the FV representation. However, before delving into specific physical models, we first derive the KG wave equation for a free relativistic scalar particle propagating in a cosmic string space-time that is both static and cylindrically symmetric. This serves as the foundation for further investigations into the behavior of particles in the presence of cosmic strings.

The line element that defines the metric of a cosmic string in a (3+1)-dimensional spacetime can be expressed in a general form as \citep{key-40,key-41,key-48,key-49,key-52}
\begin{align}
ds^{2} & =g_{\mu\nu}dx^{\mu}dx^{\nu},\nonumber \\
 & =dt^{2}-dr^{2}-\alpha^{2}r^{2}d\varphi^{2}-dz^{2},\label{eq:16}
\end{align} 
which is an exact solution to Einstein's field equations for $0\le\mu<1/4$. Furthermore, by setting $\varphi^{\prime}=\alpha\varphi$, it represents a flat conical exterior space with an angle deficit $\delta\phi=8\pi\mu$). In Eq. (\ref{eq:16}), $-\infty\le t\le+\infty$, $r\ge0$, $0\le\varphi\le2\pi$, $-\infty\le z\le+\infty,$ and $\alpha\in[0,1[$ is the parameter which determines the angular deficit $\delta\varphi=2\pi(1-\alpha)$,
and it is related to the linear mass density $\mu$ of the string by $\alpha=1-4\mu$. The line element (\ref{eq:16}) provides a mathematical representation of the spacetime geometry associated with the cosmic string, capturing the relationships between distances and intervals within this particular curved space-time. By considering the specific structure of this line element, one can examine the properties and characteristics of the cosmic string metric in a (3+1)-dimensional setting.

Alternatively, we can consider a string with a linear mass density equal to along the $z$ axis, with the Lorentz metric $ds^{2}=dt^{2}-dx'^{2}-dy'^{2}-dz'^{2}$
and the coordinate changes $x'= R \cos(\alpha\boldsymbol{\Phi}), y= R \sin(\alpha\boldsymbol{\Phi}),z=\boldsymbol{Z}$, and $t=\boldsymbol{T}$,
which leads to the line element of a cosmic string space-time with cylindrical coordinates \citep{key-m53,key-m54,key-m55,key-m56,key-m57,key-m58,key-m59}
\begin{equation}
ds^{2}=d\boldsymbol{T}^{2}-d\boldsymbol{R}^{2}-\left(\alpha\boldsymbol{R}\right)^{2}d\boldsymbol{\Phi}^{2}-d\boldsymbol{Z}^{2}.\label{eq:17}
\end{equation}
In addition to the presence of non-inertial effects, we analyze a frame
that rotates evenly with a constant angular velocity $\Omega$. The line element that describes this system is written as \citep{key-m53,key-m54,key-m55,key-m56,key-m57,key-m58,key-m59}
\begin{equation}
ds^{2}=\left(1-\alpha^{2}\Omega^{2}r^{2}\right)dt^{2}-2\Omega\alpha^{2}r^{2}d\varphi dt-dr^{2}-\alpha^{2}r^{2}d\varphi^{2}-dz^{2}.\label{eq:18}
\end{equation}
Since we do not want the term $g^{00}$ to become positive, we shall require that
\begin{equation}
0<r<\frac{1}{\alpha\Omega}.\label{eq:19}
\end{equation}
This ensures that the line element's radial coordinate (\ref{eq:18}) is defined inside the range. The metric and inverse metric tensor components are given, respectively, by \citep{key-m53,key-m54,key-m55,key-m56,key-m57,key-m58,key-m59}
\begin{equation}
g_{\mu\nu}=\left(\begin{array}{cccc}
\left(1-\left(\alpha\Omega r\right)^{2}\right) & 0 & -\Omega\left(\alpha r\right)^{2} & 0\\
0 & -1 & 0 & 0\\
-\Omega\left(\alpha r\right)^{2} & 0 & -\left(\alpha r\right)^{2} & 0\\
0 & 0 & 0 & -1
\end{array}\right),\quad g^{\mu\nu}=\left(\begin{array}{cccc}
1 & 0 & -\Omega & 0\\
0 & -1 & 0 & 0\\
-\Omega & 0 & \left(\Omega^{2}-\frac{1}{\left(\alpha r\right)^{2}}\right) & 0\\
0 & 0 & 0 & -1
\end{array}\right).\label{eq:20}
\end{equation}
It is worth noting that the problem of spinless heavy particles in
the geometry formed by a non-inertial cosmic string background has
been studied in various studies (see, for example, \citep{key-m53,key-m54,key-m55,key-m56,key-m57,key-m58,key-m59}).
To obtain the FV form of the KG wave equation in curved manifolds,
we will use the approach described in references \Citep{key-44}.
We employ the generalized Feshbach-Villars transformation (GFVT) to
describe both massive and massless particles (an equivalent transformation was presented earlier in Ref. \Citep{key-55}). The components of the wave function $\Phi$ in the GFVT are provided by \citep{key-44}
\begin{equation}
\psi=\phi_{1}+\phi_{2},\qquad i\tilde{\mathcal{D}}\psi=\mathcal{N}\left(\phi_{1}-\phi_{2}\right),\label{eq:21}
\end{equation}
where $\mathcal{N}$ is an arbitrary nonzero real parameter, and we
have defined $\tilde{\mathcal{D}}=\frac{\partial}{\partial t}+\mathcal{Y},$
with 
\begin{equation}
\mathcal{Y}=\frac{1}{2g^{00}\sqrt{-g}}\left\{ \partial_{i},\sqrt{-g}g^{0i}\right\} .\label{eq:22}
\end{equation}
The curly bracket in Eq. \eqref{eq:22} denotes the anti-commutator.
For the above-mentioned transformation, the Hamiltonian reads 
\begin{align}
\mathcal{H}_{GFVT} &= \tau_{z}\left(\frac{\mathcal{N}^{2}+\mathcal{T}}{2\mathcal{N}}\right)+i\tau_{y}\left(\frac{-\mathcal{N}^{2}+\mathcal{T}}{2\mathcal{N}}\right)-i\mathcal{Y},\nonumber \\
\mathcal{T}&= \frac{1}{g^{00}\sqrt{-g}}\partial_{i}\sqrt{-g}g^{ij}\partial_{j}+\frac{M^{2}-\xi R}{g^{00}}-\mathcal{Y}^{2}, \; \text{with}\; (i,j=1,2,3).\label{eq:23}
\end{align}
Note that for $\mathcal{N}=M$,
the original FV transformations are satisfied.

Using the metric \ref{eq:18}, we can verify that $R=0$, implying
that space-time is locally flat (there is no local gravity), and so
the coupling term is vanishing. (The condition $\xi=0$ is known as
minimum coupling. Yet, in massless theory, $\xi$ equals 1/6. (in
4 dimensions). The equations of motion are then conformally invariant in this latter instance.).
If $\mathcal{Y}\neq0,$ then, we obtain
\begin{equation}
\psi=\phi_{1}+\phi_{2},\qquad i\mathcal{D}^{\prime}\psi=\mathcal{N}\left(\phi_{1}-\phi_{2}\right),\label{eq:24}
\end{equation}
with $\mathcal{D}^{\prime}=\frac{\partial}{\partial t}+\mathcal{Y^{\prime}}.$
The next step is to extend the method employed in Sec. \ref{sec2} to the situation of spinning cosmic strings. The GFVT is used to construct the equations of motion for this issue, which are then solved to give the wave functions and energy spectra. According to the Refs. \Citep{key-52,key-53}, identifying the Hamiltonian leads to the concept of using a two-component formulation of the KG-type fields. To recast Eq. \eqref{eq:26} in the Hamiltonian form, additional definitions for the quantities described in Eqs. \eqref{eq:20} must be introduced Eqs. \eqref{eq:22} to \eqref{eq:30} \Citep{key-54}.
In this way, we write
\begin{equation}
\mathcal{H}_{GFVT}^{\prime}=\tau_{z}\left(\frac{\mathcal{N}^{2}+\mathcal{T^{\prime}}}{2\mathcal{N}}\right)+i\tau_{y}\left(\frac{-\mathcal{N}^{2}+\mathcal{T^{\prime}}}{2\mathcal{N}}\right)-i\mathcal{Y}^{\prime},\label{eq:25}
\end{equation}
with
\begin{align}
\mathcal{T}^{\prime} & =\partial_{i}\frac{G^{ij}}{g^{00}}\partial_{j}+\frac{M^{2}-\xi R}{g^{00}}+\frac{1}{\mathcal{F}}\nabla_{i}\left(\sqrt{-g}G^{ij}\right)\nabla_{j}\left(\frac{1}{\mathcal{F}}\right)+\sqrt{\frac{\sqrt{-g}}{g^{00}}}G^{ij}\nabla_{i}\nabla_{j}\left(\frac{1}{\mathcal{F}}\right)+\frac{1}{4\mathcal{F}^{4}}\left[\nabla_{i}\left(\mathcal{U}^{i}\right)\right]^{2}\nonumber \\
 & -\frac{1}{2\mathcal{F}^{2}}\nabla_{i}\left(\frac{g^{0i}}{g^{00}}\right)\nabla_{j}\left(\mathcal{U}^{i}\right)-\frac{g^{0i}}{2g^{00}\mathcal{F}^{2}}\nabla_{i}\nabla_{j}\left(\mathcal{U}^{i}\right),\label{eq:26}
\end{align}
and
\begin{equation}
\mathcal{Y}^{\prime}=\frac{1}{2}\left\{ \partial_{i},\sqrt{-g}\frac{g^{0i}}{g^{00}}\right\}, \label{eq:27}
\end{equation}
where
\begin{equation}
G^{ij}=g^{ij}-\frac{g^{0i}g^{0j}}{g^{00}},\;\;\mathcal{F}=\sqrt{g^{00}\sqrt{-g}},\;\;\mathcal{U}^{i}=\sqrt{-g}g^{0i}.\label{eq:28}
\end{equation}
Reference \Citep{key-54} demonstrates that the transformations
\eqref{eq:25}, \eqref{eq:26}, and \eqref{eq:27} with the definitions in Eq. \eqref{eq:28}
are precise and encompass all inertial and gravitational fields. It
should be emphasized that using these precise transformations guarantees
that the block-diagonal form of the Hamiltonian $\mathcal{H}_{GFVT}$
is obtained, which is independent of the parameter $\mathcal{N}$. The eigenvalues
of the operator $\mathcal{Y}^{\prime}$ for the geometry in Eq. \eqref{eq:18}
are given by
\begin{equation}
\mathcal{Y}^{\prime}\Psi\left(t,r,\varphi\right)=\frac{1}{2}\left\{ \partial_{2},\sqrt{-g}\frac{g^{02}}{g^{00}}\right\} \Psi\left(t,r,\varphi\right)=\frac{g^{02}}{g^{00}}\frac{\partial}{\partial\varphi}\Psi\left(t,r,\varphi\right),\label{eq:29}
\end{equation}
where the field $\Psi$ obeys the non-unitary transformation $\Psi\equiv\Phi^{\prime}=\mathcal{F}\Phi$,
which permits to obtain pseudo-Hermitian Hamiltonian $\mathcal{H}_{GFVT}^{\prime}=\mathcal{F}\mathcal{H}_{GFVT}^{\prime}\mathcal{F}^{-1}$, with $\mathcal{H}_{GFVT}^{\prime}=\tau_{z}\left(\mathcal{H}_{GFVT}^{\prime}\right)^{\dagger}\tau_{z}.$
After some mathematical calculations, Eq. \eqref{eq:26} takes the form
\begin{equation}
\mathcal{T}^{\prime}=-\frac{1}{\mathcal{F}}\left[\left(\frac{\partial}{\partial r}\right)\left(\sqrt{-g}\frac{\partial}{\partial r}\right)+\sqrt{-g}\left(\varOmega^{2}-\frac{1}{\alpha^{2}r^{2}}\right)\frac{\partial^{2}}{\partial\varphi^{2}}+\sqrt{-g}\frac{\partial^{2}}{\partial z^{2}}\right]\frac{1}{\mathcal{F}}+\frac{M^{2}}{g^{00}}-\left(\frac{g^{02}}{g^{00}}\frac{\partial}{\partial\varphi}\right)^{2},\label{eq:30}
\end{equation}
where for the metric \eqref{eq:18} the Ricci scalar vanishes i.e,
$R=0$. The goal of using the Hamiltonian \eqref{eq:25} and then using the Hamiltonian \eqref{eq:28} is to find solutions in the form of the Schrödinger equation, such as
\begin{equation}
\mathcal{H}_{GFVT}^{\prime}\Psi^{\prime}(\boldsymbol{x})=i\,\frac{d}{dt}\Psi^{\prime}(\boldsymbol{x}),\qquad\boldsymbol{x}\equiv(t,r,\varphi,z).\label{eq:31}
\end{equation}
Consider the ansatz below for the wave function to solve this eigenvalue
problem
\begin{equation}
\Psi(\boldsymbol{x})=\mathcal{F}\Phi\left(\boldsymbol{x}\right)=\mathcal{F}\left(\begin{array}{c}
\phi_{1}\left(\boldsymbol{r}\right)\\
\phi_{2}\left(\boldsymbol{r}\right)
\end{array}\right)e^{-\left(iEt-j\varphi-k_{z} z\right)},\label{eq:32}
\end{equation}
where $j=0,\pm 1,\pm 2,\pm 3,\ldots $ is the angular momentum quantum number, and $k_{z}\in \left[ -\infty ,\infty \right] $. By replacing Eq. \eqref{eq:32} into  Eq. \eqref{eq:31}, we obtain the differential equations 
\begin{equation} 
\begin{aligned}\mathcal{N}^{2}\mathcal{F}\left(\phi_{1}-\phi_{2}\right)+\mathcal{T}^{\prime}\mathcal{F}\left(\phi_{1}-\phi_{2}\right)+2\mathcal{N}\left(\frac{g^{02}}{g^{00}}j\right)\mathcal{F}\phi_{1} & =2\mathcal{N}E\mathcal{F}\phi_{1},\\
\mathcal{N}^{2}\mathcal{F}\left(\phi_{1}-\phi_{2}\right)-\mathcal{T}^{\prime}\mathcal{F}\left(\phi_{1}+\phi_{2}\right)+2\mathcal{N}\left(\frac{g^{02}}{g^{00}}j\right)\mathcal{F}\phi_{2} & =2\mathcal{N}E\mathcal{F}\phi_{2}.
\end{aligned}
\label{eq:33}
\end{equation}
Adding and subtracting these equations leads, after simplification, to sets of coupled equations for $\phi_{1}$ and $\phi_{2}$
\begin{align}
&\mathcal{N}\mathcal{F}\left(\phi_{1}-\phi_{2}\right)+\left(\frac{g^{02}}{g^{00}}j\right)\mathcal{F}\left(\phi_{1}+\phi_{2}\right) =E\mathcal{F}\left(\phi_{1}+\phi_{2}\right),\label{eq:34}\\
-&\mathcal{T}^{\prime}\mathcal{F}\left(\phi_{1}+\phi_{2}\right)+\mathcal{N}\left(\frac{g^{02}}{g^{00}}j\right)\mathcal{F}\left(\phi_{1}-\phi_{2}\right)=\mathcal{N}E\mathcal{F}\left(\phi_{1}-\phi_{2}\right),\label{eq:35}
\end{align}
where we have used the result
\begin{equation}
\phi_{1}=  \frac{1}{\mathcal{N}}\left(E-\frac{g^{02}}{g^{00}}j\right)\phi_{2}
=  \frac{1}{\mathcal{N}}\left(E+\varOmega j\right)\phi_{2}.\label{eq:36}
\end{equation}
After some simple algebraic manipulations, we arrive at the following second-order differential equation for the radial function $\psi(r)$:
\begin{equation}
\left[\frac{d^{2}}{dr^{2}}+\frac{1}{r}\frac{d}{dr}-\frac{\varsigma^{2}}{r^{2}}+\gamma^{2}\right]\varphi_{1}\left(r\right)=0, \label{eq:38}
\end{equation}
where we have defined the parameters $\varsigma^{2}=j^{2}/\alpha^{2}$ and $\gamma^{2}=\left(E+\varOmega j\right)^{2}-M^{2}-k_{z}^{2}$. It can be shown that Eq. \eqref{eq:38} is a Bessel-type differential equation, and its general solution is given by
\begin{equation}
\psi \left( r\right) =A^{\prime }\,J_{\varsigma }\left( \gamma r\right)
+B^{\prime }N_{\varsigma }\left( \gamma r\right) ,  \label{gs}
\end{equation}
where $J_{\varsigma }\left( \gamma r\right) $ and $N_{\varsigma }\left(
\gamma r\right) $ are the Bessel function of the first and second kind, and $
A^{\prime }$ and $B^{\prime }$ are arbitrary constants. In the next sections, we make some applications of the above formalism. We shall consider models already consolidated in the literature to make clear the physical implications due to the non-inertial effects and the curvature parameter of the cosmic string.

\section{Applications}\label{sec4}

\subsection{Particle confined to a hard-wall potential}

Studying the relativistic motion of a spin-0 particle confined to a hard-wall potential is important in theoretical physics and provides us with very interesting results. By exploring this scenario, we gain valuable insights into the behavior of particles under extreme conditions and the interplay between relativity and quantum mechanics. This study lets us deepen our understanding of fundamental principles, refine theoretical frameworks, and develop computational techniques. Moreover, it serves as a stepping stone for comprehending more complex systems, guiding us toward a more complete understanding of the universe. Through this investigation, we unveil the intricate dynamics of spin-0 particles and contribute to the broader advancement of physics as a whole. For the simple model described by the solution of Eq. (\ref{eq:38}), we must consider the physically acceptable solution in the region of interest. In this case, since solution (\ref{gs}) must be regular at the origin and function $N_{\varsigma }\left( \gamma r\right) $ is not well behaved in this region, we must make $B=0$ in the solution (\ref{gs}). The relevant solution is 
\begin{equation}
\psi \left( r\right) =A^{\prime }\,J_{\varsigma }\left( \gamma r\right).
\label{eq:39}
\end{equation}
\begin{figure}[!h]
\centering
\includegraphics[scale=0.35]{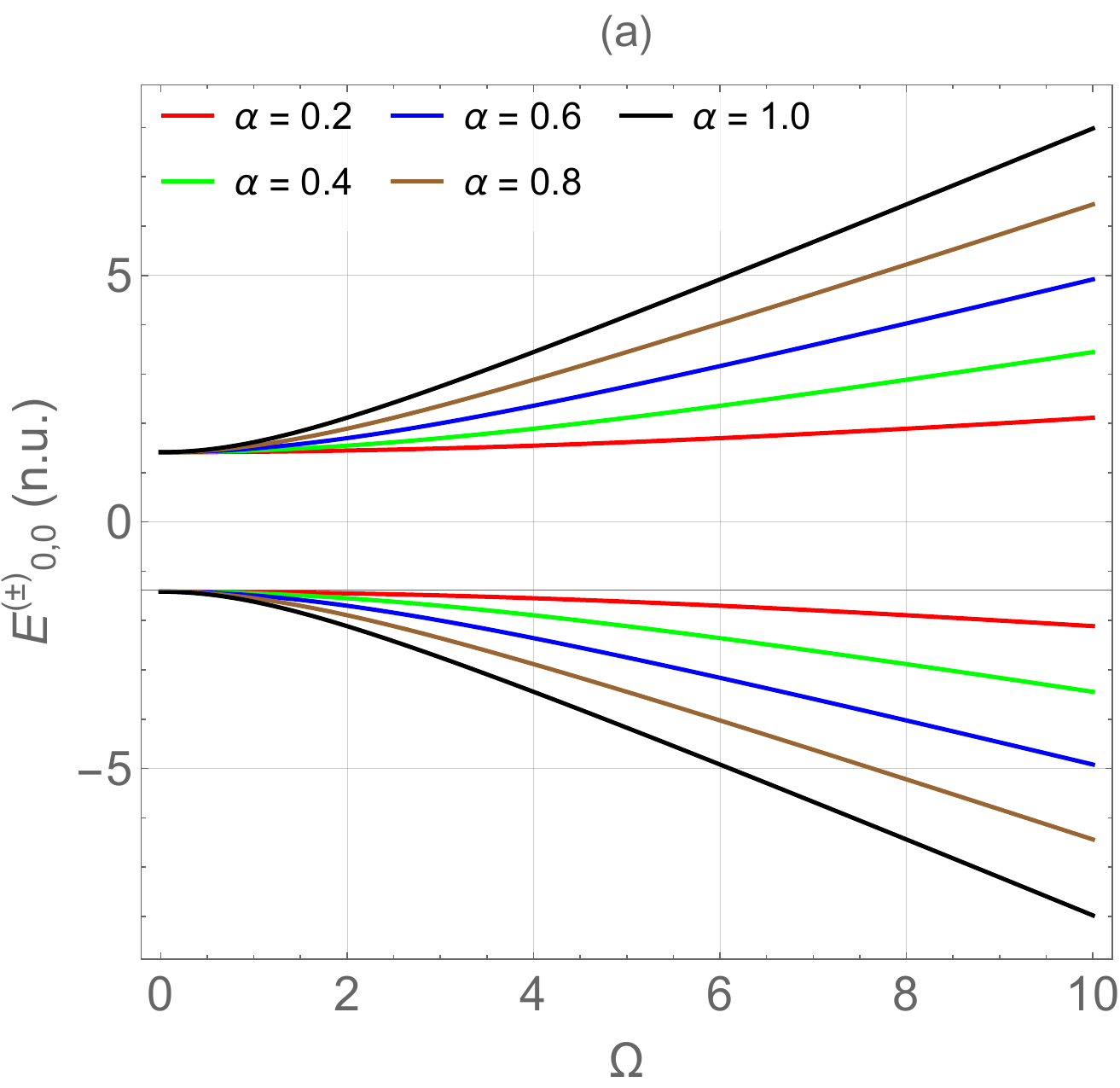}
\includegraphics[scale=0.35]{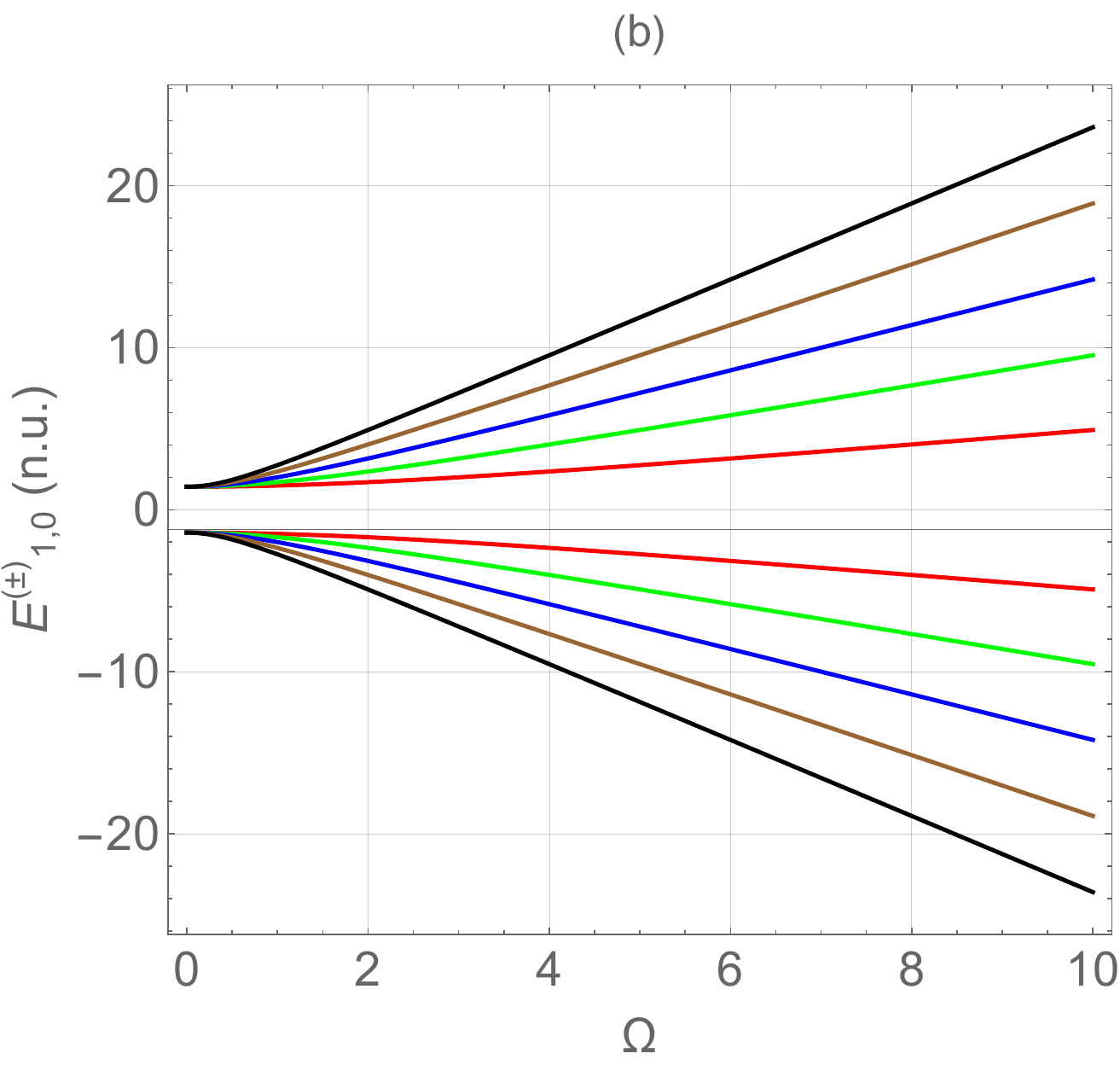}
\includegraphics[scale=0.35]{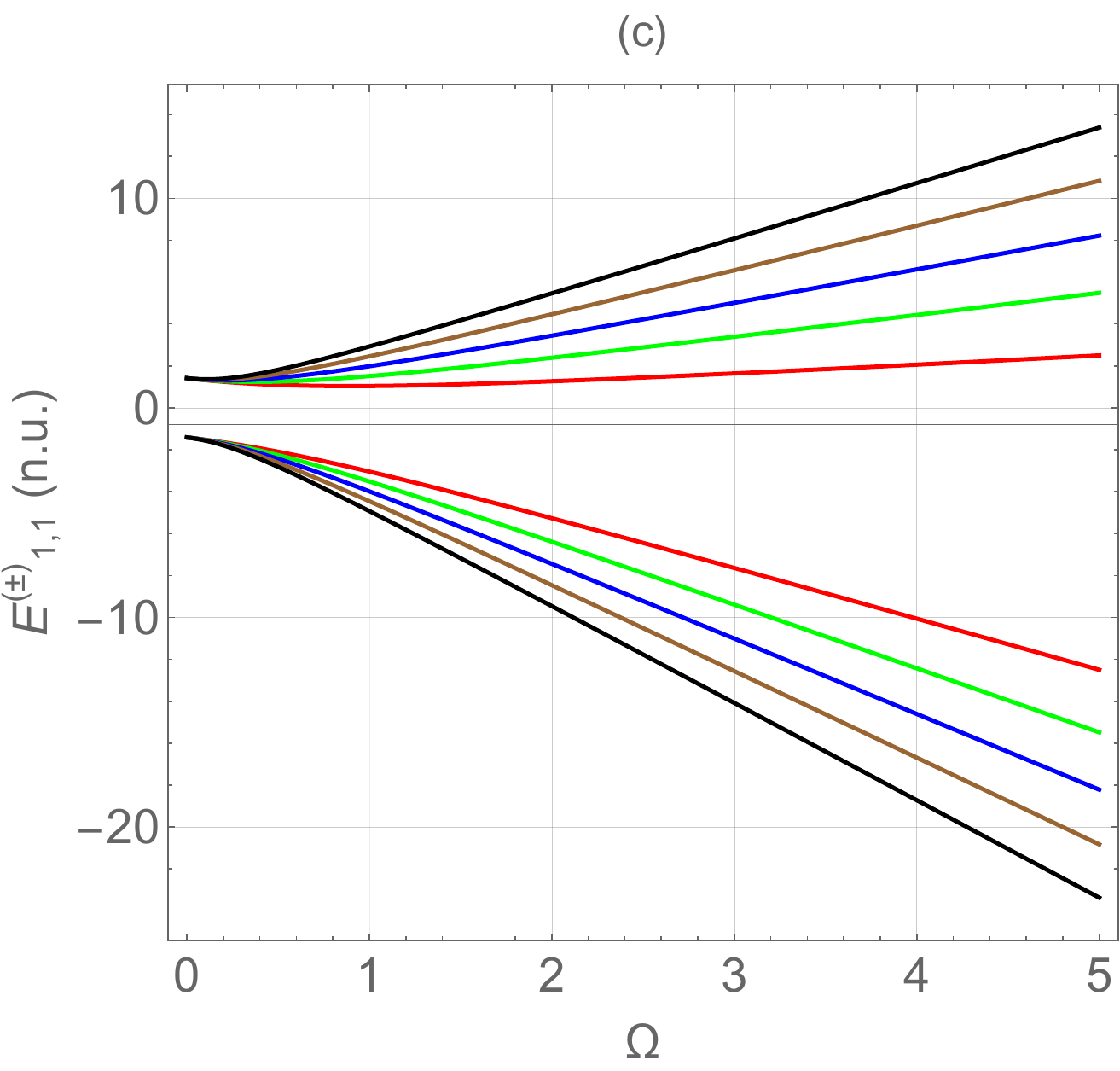}
\includegraphics[scale=0.35]{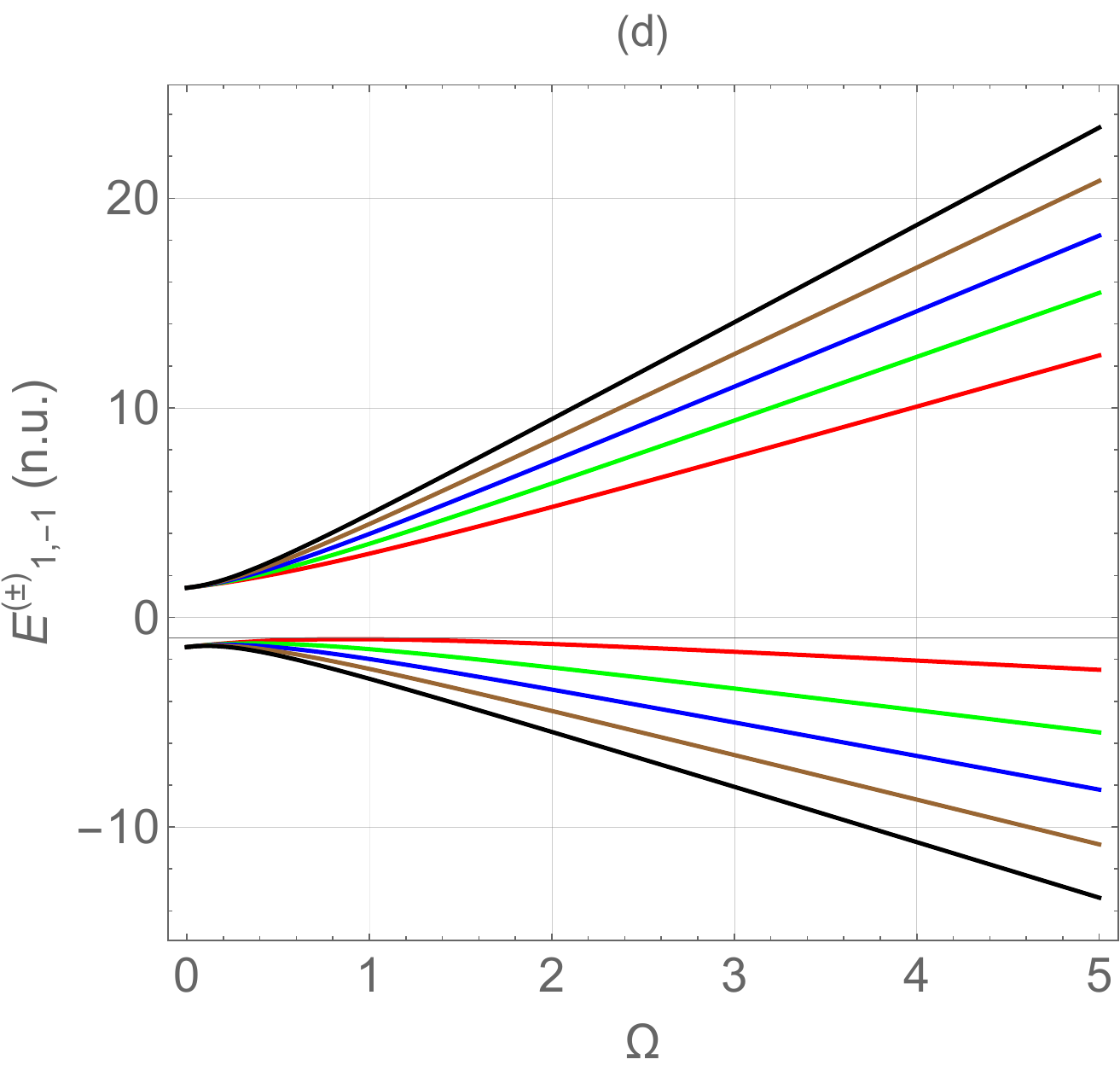}
\caption{Sketch of the energies (Eq. (\ref{bessel_energies})) as a function of $\Omega$ for different values of $n$ and $\alpha$. In (a), the profile for $n=0$ and $j=0$, (b) $n=1$ and $j=0$, (c) $n=1$ and $j=1$, and (d) $n=1$ and $j=-1$. We use $M=1$, $k_{z}=1$, and $\kappa=2.5$. A manifestation due to curvature effects is observed when $\alpha$ is decreased, which implies $|E_{n,j}^{(\pm})|$ increasing.}
\label{FigBessel}	
\end{figure}
To have a normalized wave function in the region established by the range
\eqref{eq:19}, we must require that
\begin{equation}
\psi \left( r\rightarrow r_{0}=\frac{1}{\alpha \Omega }\right) =0.
\label{cnd}
\end{equation}
This condition remains valid for $r_{0}\gg 1$ and fixed values of $\varsigma 
$ and $\gamma $. In this limit, we can write the solution (\ref{eq:39}) in
the form 
\begin{equation}
\psi \left( \boldsymbol{r}\right) =|\mathcal{C}_{1}|\sqrt{\frac{2}{\pi
\gamma r_{0}}}\left( 
\begin{array}{c}
1+\frac{E}{\mathcal{N}} \\ 
1-\frac{E}{\mathcal{N}}
\end{array}
\right) e^{-i\left( Et-j\varphi -k_{z}z\right) }\,\cos \left( \gamma r_{0}-\frac{
\varsigma \pi }{2}-\frac{\pi }{4}\right).  \label{eq:40}
\end{equation}
By applying condition (\ref{cnd}) to the solution (\ref{eq:40}), we obtain
the relation
\begin{equation}
\gamma r_{0}-\frac{\varsigma \pi }{2}-\frac{\pi }{4}=\frac{n\pi }{2},\text{with} \,n \in \mathbb{Z^{*}},
\end{equation}
which solved for $E$ provides
\begin{equation}
E_{n,j}^{(\pm)} =-j\Omega \pm \sqrt{\left( \frac{1}{2}\pi \Omega \left\vert j\right\vert +
\frac{1}{4}\pi \Omega \alpha +\frac{1}{2}n\pi \Omega \alpha \right)\label{bessel_energies}
^{2}+k_{z}^{2}+M^{2}},
\end{equation}
which represents the particle's energy. By studying these energies, we can report some inherent characteristics of the model. We evaluate some particular states of the energy (\ref{bessel_energies}) as a function of $\Omega$. All graphical analyses in our article we use $M=1$, $k_{z}=1$, and $\kappa=2.5$ (Fig. \ref{FigBessel}). We can see that $E_{n,j}$ exhibits different profiles when $\alpha$ changes, for example, when $\alpha=0.2, 0.4, 0.6, 0.8, 1.0$, $|E_{0,1}^{(\pm)}|$ is symmetric about $\Omega=0$, and increases when $\Omega$ is increased. When this occurs, it is observed that the spectrum becomes asymmetric (Fig. \ref{FigBessel}(a)). These features are also observed when we investigate $E_{1,0}$, but with the difference that $|E_{0,1}^{(\pm)}|<|E_{1,0}^{(\pm)}|$ (Fig. \ref{FigBessel}(b)).  An interesting feature is observed when we study the energies for $n=1,j=1$ (Fig. \ref{FigBessel}(c)) and $n=1,j=-1$ (Fig. \ref{FigBessel}(d)). We observe that $|E_{1,1}^{(\pm)}|$ and $|E_{1,-1}^{(\pm)}|$ increase with $\Omega$. However, $|E_{1,1}^{(+)}|< |E_{1,-1}^{(+)}|$ while $|E_{1,1}^{(-)}|>|E_{1,-1}^{(-)}|$. This manifestation is due to the combined effects of rotation and the parameter $\alpha$. 

\subsection{The Coulomb-type potential}

In this section, we solve the KG equation with a Coulomb-like potential in the FV formalism. Including this potential in the KG equation allows investigating of long-range electromagnetic interactions and describing the interaction between charged particles. This investigation is crucial for understanding various physical phenomena, from atomic structure to elementary particle physics. We must consider an attractive Coulomb-type potential to solve the equation then and analyze fundamental properties such as bound states. We include the potential in the equation of motion by substituting $E \rightarrow E-V(r)$ into Eq. (\ref{eq:38}). The potential $V(r)$ is specified by\begin{equation}
V\left(r\right)=-\frac{\kappa}{r},\label{eq:41}
\end{equation}
where $\kappa$ is a positive arbitrary parameter.
The radial equation to be solved is
\begin{equation}
\left[\frac{d^{2}}{dr^{2}}+\frac{1}{r}\frac{d}{d r}-\frac{\zeta^{2}}{r^{2}}+\frac{2\lambda}{r}-\gamma^{2}\right]\varphi\left(r\right)=0,\label{eq:45}
\end{equation}
where \ensuremath{\zeta^{2}=\varsigma^{2}+\kappa^{2},\gamma^{2}=M^{2}+k_{z}^{2}-\left(\mathcal{E}+\varOmega j\right)^{2}}, and $\lambda=\kappa\left(\mathcal{E}+\varOmega j\right)$. The radial equation (\ref{eq:45}) is of the confluent hypergeometric type, and a solution satisfying the asymptotic limits for both $r \rightarrow 0$ and $r \rightarrow \infty$ can be found. Furthermore, this equation appears in several problems solved in the literature, so we do not find it necessary to present a detailed solution. Thus, using the results above, it can be shown that the eigenfunctions and energy eigenvalues of Eq. (\ref{eq:45}) in the FV representation are respectively given by
\begin{equation}
\psi\left(\boldsymbol{r}\right)=C_n \left(\begin{array}{c}
1+\frac{E}{\mathcal{N}}\\
1-\frac{E}{\mathcal{N}}
\end{array}\right)e^{-i\left(Et-j\zeta-k_{z} z\right)}\,(2 \gamma )^{\zeta+\frac{1}{2}} e^{-\gamma  r} r^{\zeta} \, _1F_1\left(-n,2 \zeta+1,2 \gamma  r\right),\label{coulomb_eigenf}
\end{equation}
\begin{equation}
\mathcal{E}^{(\pm)}_{n,j}=-j\Omega\pm\frac{(2\zeta+2n+1)^{2}\left(k_{z}^{2}+M^{2}\right)}{\sqrt{(2\zeta+2n+1)^{2}\left(4\kappa^{2}+(2\zeta+2n+1)^{2}\right)\left(k_{z}^{2}+M^{2}\right)}}.\label{encl}
\end{equation}
\begin{figure}[!t]
\centering
\includegraphics[scale=0.35]{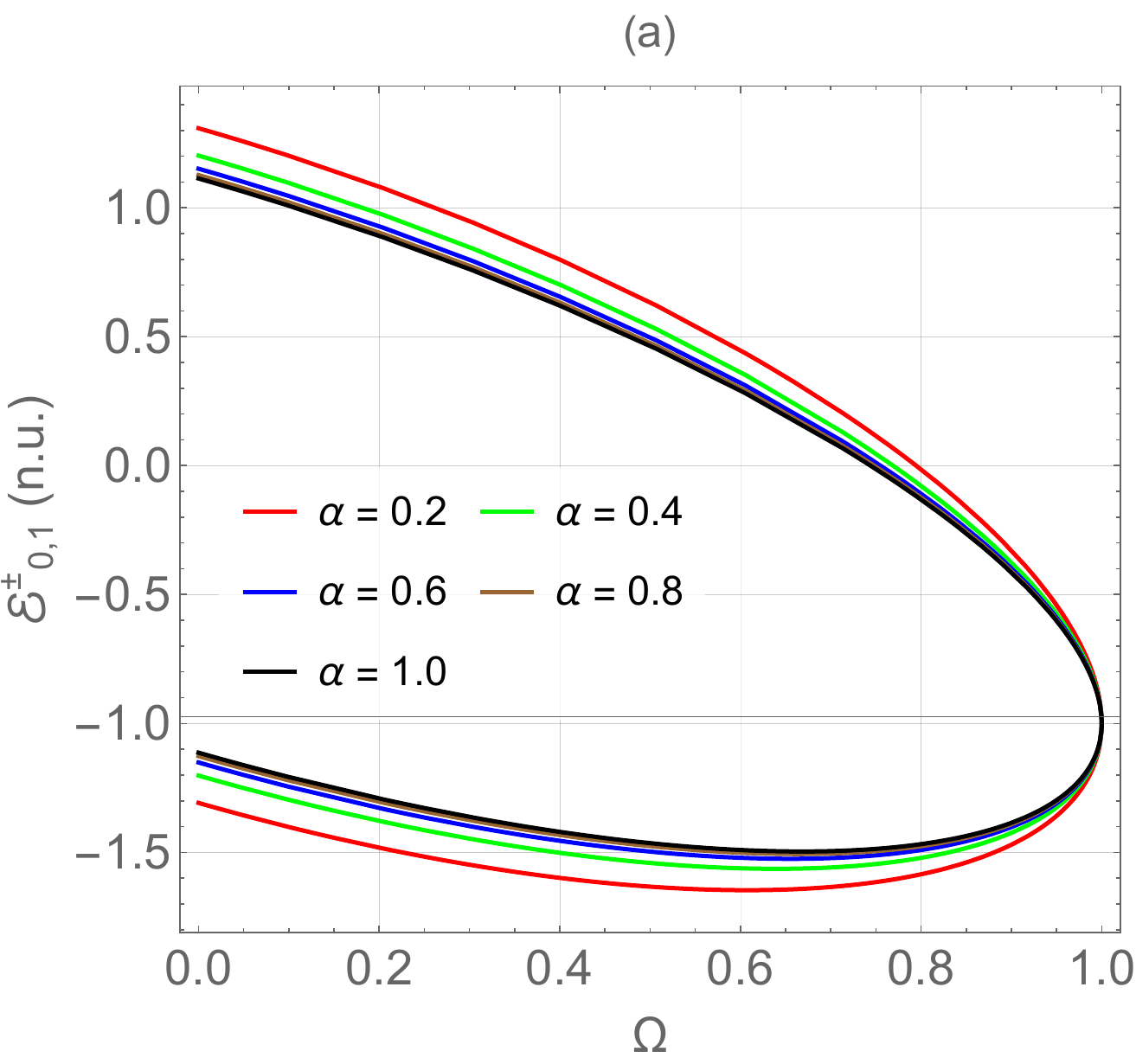}
\includegraphics[scale=0.35]{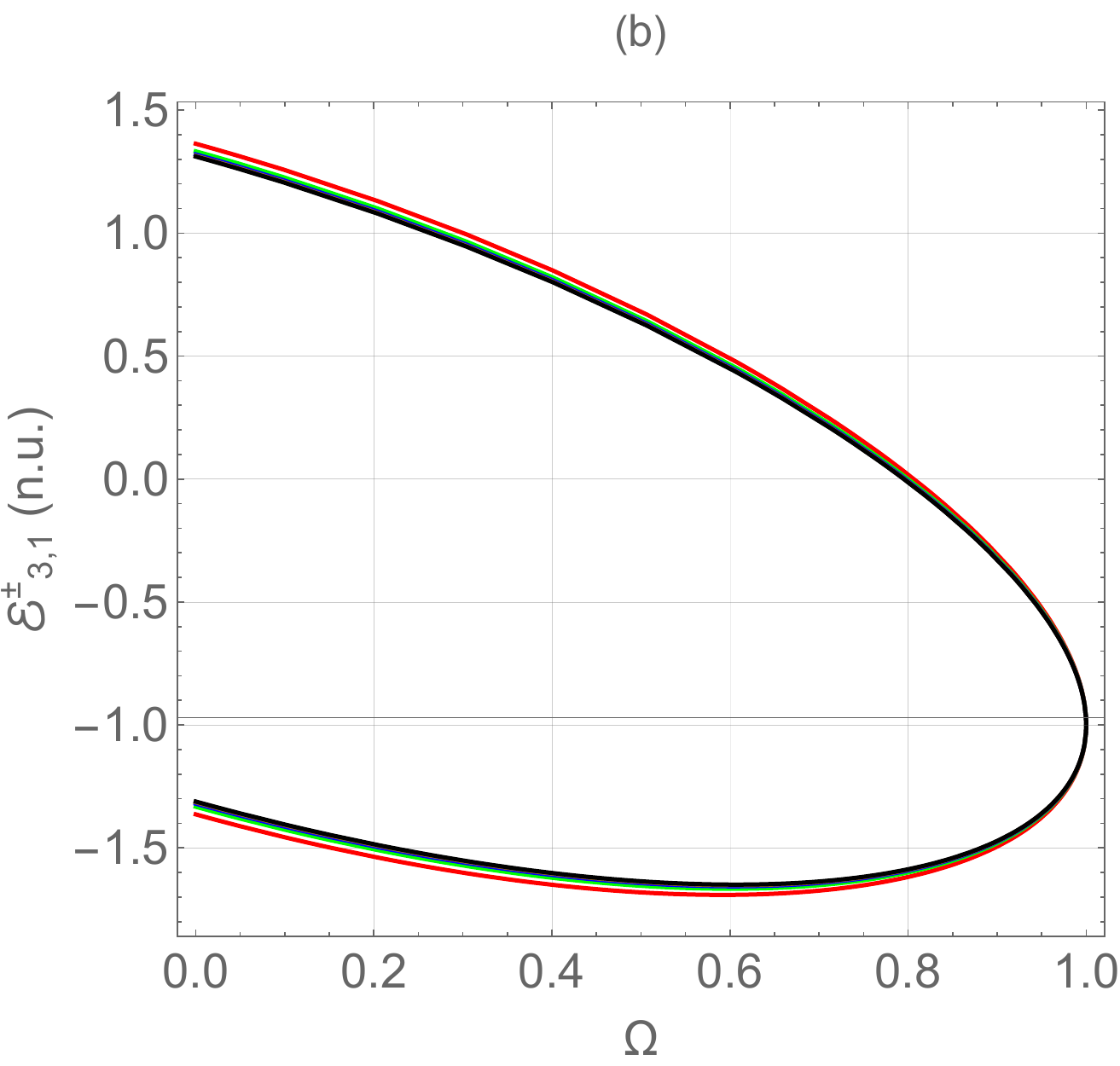}
\includegraphics[scale=0.35]{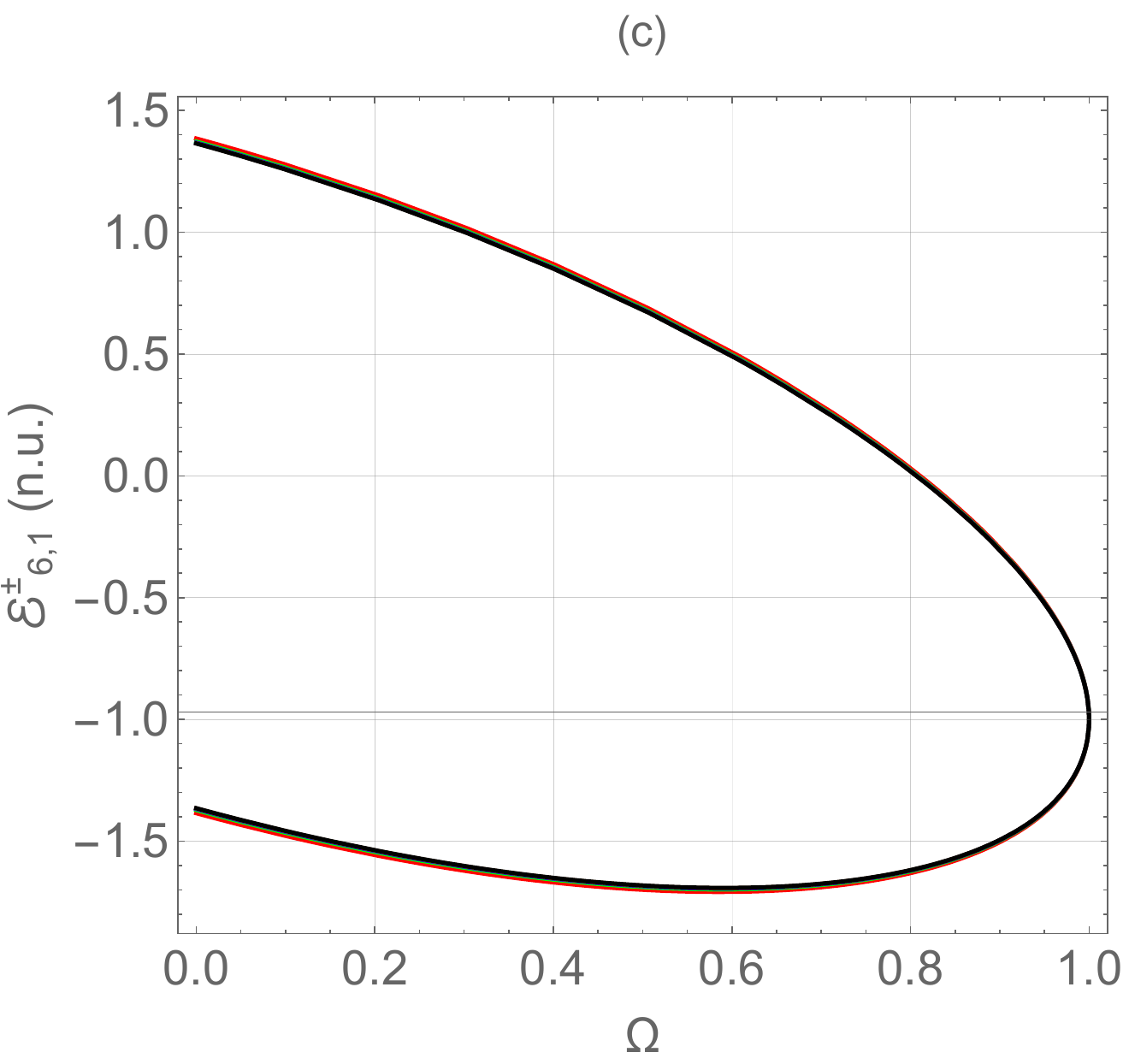}
\includegraphics[scale=0.35]{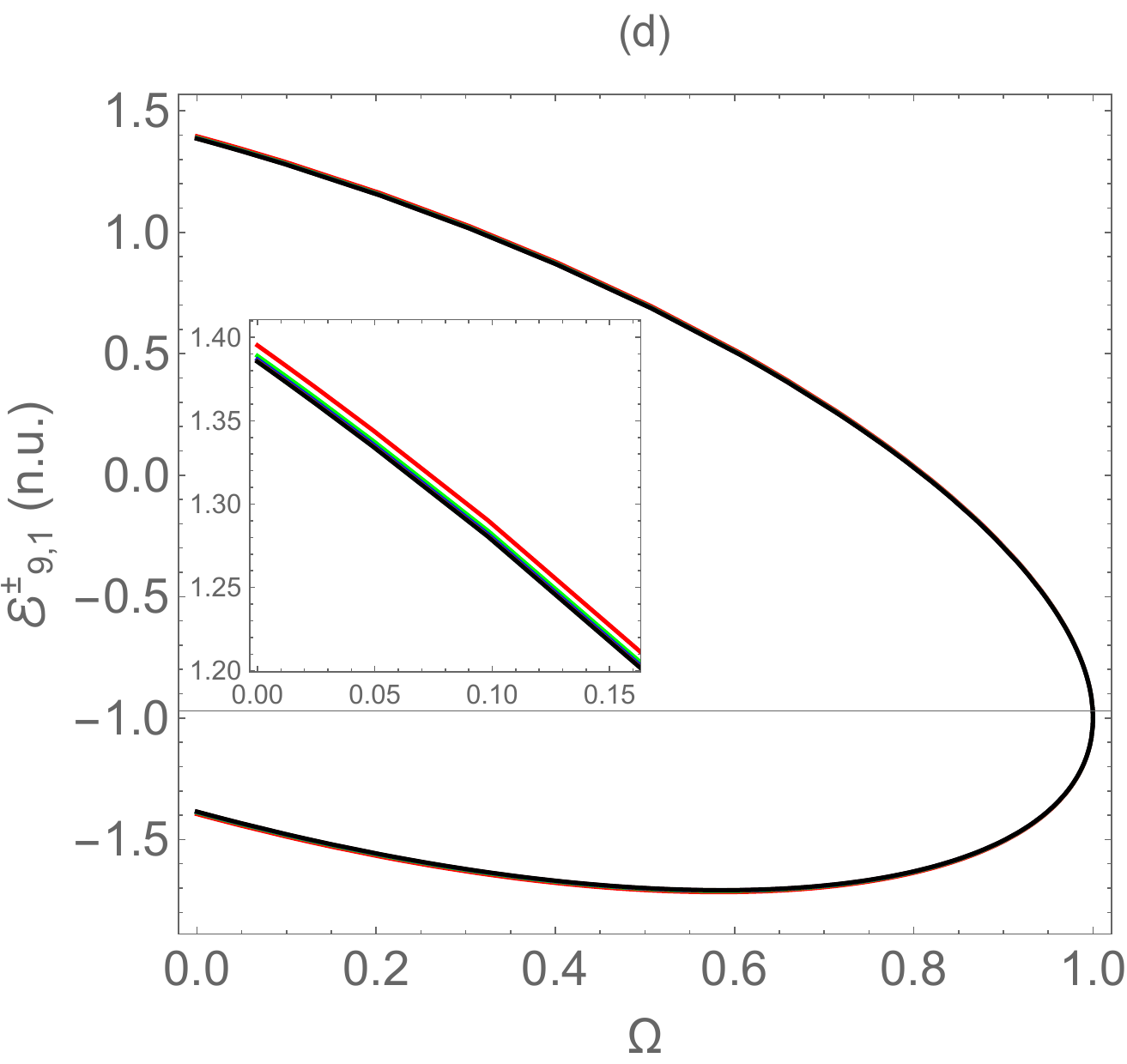}
\caption{Sketch of the energies (Eq. (\ref{encl})) as a function of $\Omega$ for different values of $n$ and $\alpha$. In (a), the profile for $n=0$ and $j=1$, (b) $n=3$ and $j=1$, (c) $n=6$ and $j=1$,and (d) $n=9$ and $j=1$. We use $M=1$, $k_{z}=1$, and $\kappa=2.5$. The states with increasing $n$ reveal an approximation between the energy curves. When $\alpha$ is decreased, $|E_{n,j}^{(\pm})|$ increases.}
\label{FigCoulomb}	
\end{figure}
\begin{figure}[!t]
\centering
\includegraphics[scale=0.35]{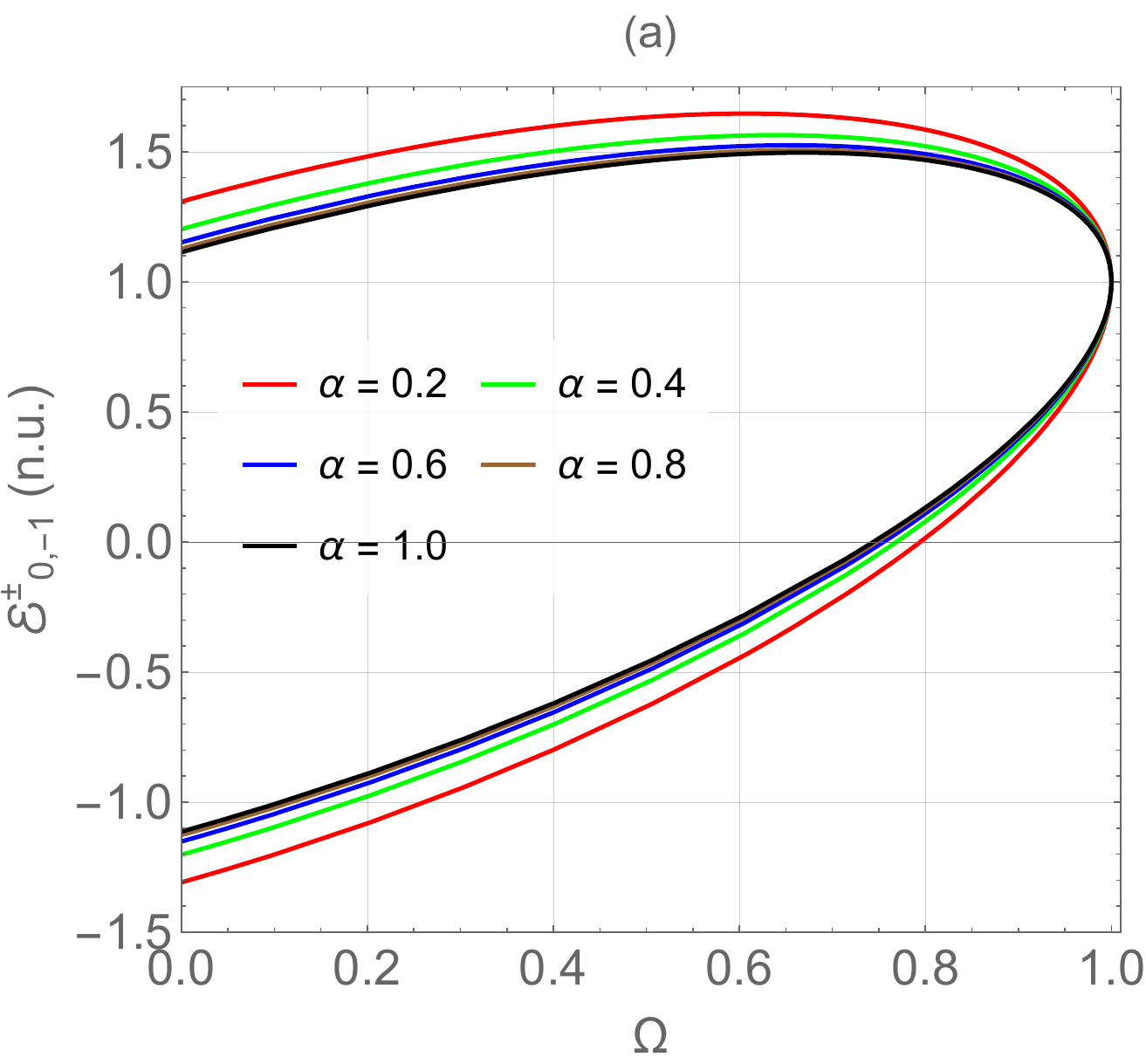}
\includegraphics[scale=0.35]{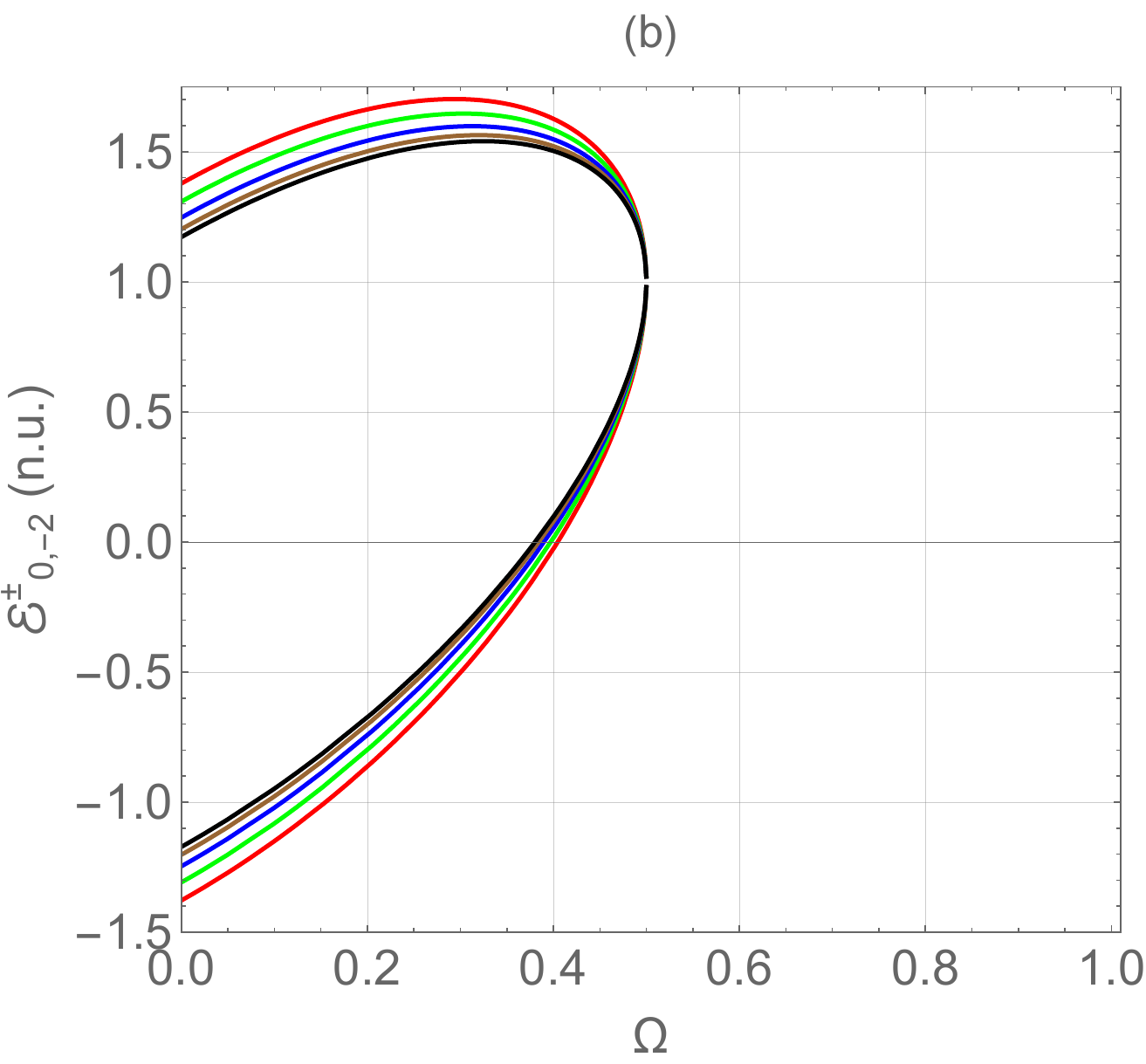}
\includegraphics[scale=0.35]{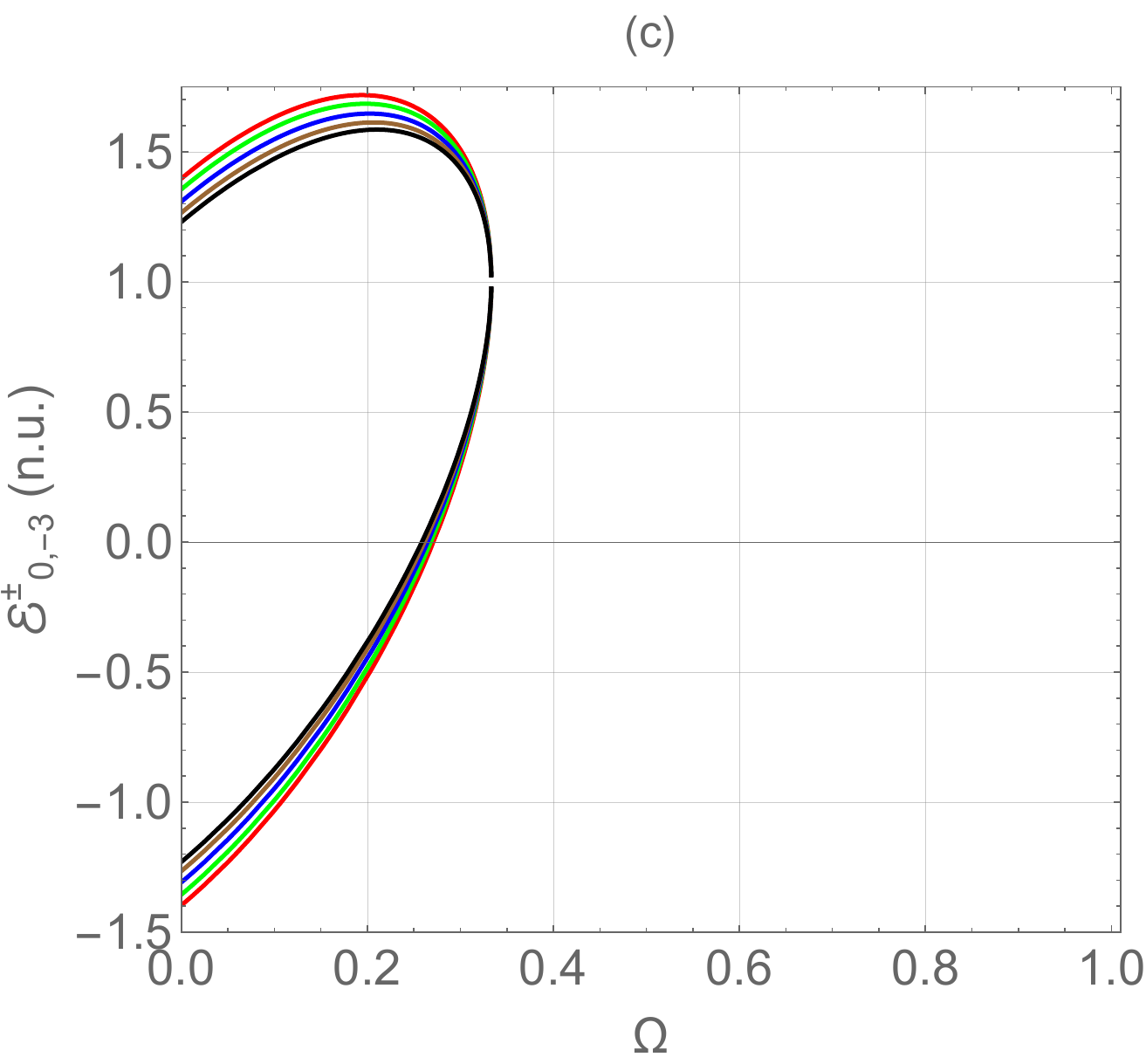}
\includegraphics[scale=0.35]{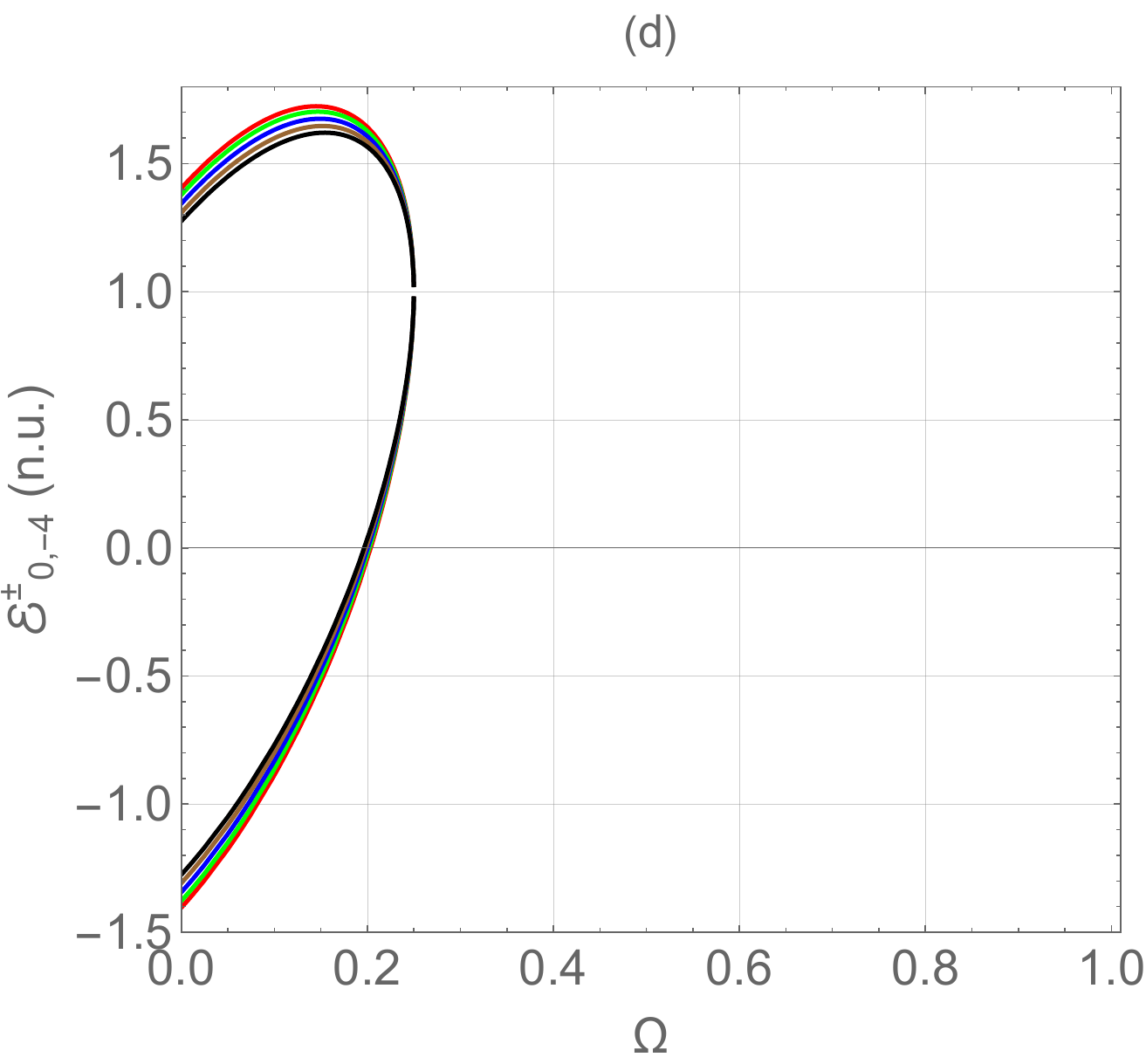}
\caption{Sketch of the energies (Eq. (\ref{encl})) as a function of $\Omega$ for different values of $n$ and $\alpha$. In (a), the profile for $n=0$ and $j=-1$, (b) $n=0$ and $j=-3$, (c) $n=0$ and $j=-5$; and (d) $n=0$ and $j=-7$. We use $M=1$, $k_{z}=1$, and $\kappa=2.5$. In addition to the features present in Fig. \ref{FigCoulomb}, when we set $n$ and consider different negative values of $m$, the energies are defined in a rotation range that tends to decrease with negative $m$.}
\label{FigCoulombm}	
\end{figure}
Possible choices for the parameters in (\ref{encl}) can display different profiles. For example, analyzing the states with $j=1$ and $n=0,3,6,9$, the positive and negative values of $\mathcal{E}^{(\pm)}_{n,j}$ are shown in Fig. \ref{FigCoulomb}. The curvature effects on the energies directly impact the spacing between the energy curves, making this spacing somewhat smaller for increasingly larger values of $n$ (Figs. \ref{FigCoulomb}(a)-(d)). The inset in Fig. \ref{FigCoulomb}(d) shows that such spacings are maintained, and the energy has an approximately linear change over small rotation intervals. For this particular configuration, $|\mathcal{E}^{(\pm)}_{n,j}|$ decreases when $\Omega$ increases, and it exists only in the interval $0<\Omega<1$. A second configuration is displayed in Fig. \ref{FigCoulombm} for $n=0$ and $j=-1,-2,-3,-4$. For this particular configuration, $|\mathcal{E}_{nj}^{\pm}|$ increases when $\alpha$ is reduced. However, when $m$ takes on larger values, the range of $\Omega$ that defines the allowed energies is reduced. This effect becomes clear when we look at the energy profiles in Figs. \ref{FigCoulombm}(a)-\ref{FigCoulombm}(d).

\subsection{The Feshbach-Villars Oscillator (FVO)}\label{secKGO}

The final application of our study involves including the FVO in the equation of motion. The FVO is a well-known, exactly solvable model in relativistic quantum mechanics that have been constructed in both flat and curved spacetime \Citep{key-48}. This model combines the principles of special relativity with quantum mechanics, providing insights into the dynamics of spin-0 particles in a harmonic potential. By exploring the FVO, we gain a deeper understanding of wave equations, energy spectra, and the quantization of fields. Moreover, this study opens doors to investigating phenomena like particle creation and annihilation, vacuum fluctuations, and the manifestation of symmetries. As an extension, it is important to study the FVO in other physical contexts because it helps us to expand our knowledge of fundamental physics, improve our analytical capabilities to solve new equations of motion, and prepare the way for future advances in quantum field theory. 
The FVO is introduced into equation \eqref{eq:26} and \eqref{eq:27} by substituting $\partial_{r}\rightarrow \partial_{r}+M\omega r$. So, we write
\begin{align}
\mathcal{T}^{\prime}&=-\frac{1}{\mathcal{F}}\left[\left(\frac{\partial}{\partial r}+M\omega r\right)\left(\sqrt{-g}\left(\frac{\partial}{\partial r}-M \omega r\right)\right)+\sqrt{-g}\left(\varOmega^{2}-\frac{1}{\alpha^{2}r^{2}}\right)\frac{\partial^{2}}{\partial\varphi^{2}}+\sqrt{-g}\frac{\partial^{2}}{\partial z^{2}}\right]\frac{1}{\mathcal{F}}\notag \\&+\frac{M^{2}}{g^{00}}-\left(\frac{g^{02}}{g^{00}}\frac{\partial}{\partial\varphi}\right)^{2}.\label{eq:52}
\end{align}
\begin{equation}
    \mathcal{Y}^{\prime} =0.
    \label{eq:52.1}
\end{equation}
Equation \eqref{eq:52} can be solved by following the same steps as before. We shall insert Eqs. \eqref{eq:52} and \eqref{eq:29} into the Hamiltonian
\eqref{eq:25}. Subsequently, we assume the solution \eqref{eq:32} gives two
coupled differential equations similar to those of Eq. \eqref{eq:33},
but with different values of $\mathcal{T}^{\prime}$. We obtain the radial equation
\begin{equation}
\left[\frac{d^{2}}{dr^{2}}+\frac{1}{r}\frac{d}{dr}-M^{2}\omega^{2}r^{2}-\frac{\vartheta^{2}}{r^{2}}+\delta\right]\psi\left(r\right)=0,\label{eq:53}
\end{equation}
where we have defined the parameters
\begin{equation}
\vartheta^{2}=\frac{j^{2}}{\alpha^{2}},\;\;\delta=\left(E+\varOmega j\right)^{2}-M^{2}-k_{z}^{2}+M\omega.\label{eq:54}
\end{equation}
To solve Eq. \eqref{eq:53}, we introduce the new dimensionless variable
$\mathcal{U}=M\omega r^{2}$, and replace it in Eq. \ref{eq:53}. The resulting equation reads
\begin{equation}
\left[\frac{d^{2}}{d\mathcal{U}^{2}}+\frac{1}{\mathcal{U}}\frac{d}{d\mathcal{U}}-\frac{\vartheta^{2}}{4\mathcal{U}^{2}}-\frac{1}{4}+\frac{\delta}{4M\omega\mathcal{U}}\right]\psi\left(\mathcal{U}\right)=0.\label{eq:55}
\end{equation}
Now, we introduce the new function 
\begin{equation}
\psi(\mathcal{U})=\mathcal{U}^{-\frac{1}{2}}\mathcal{J}\left(\mathcal{U}\right).\label{eq:56}
\end{equation}
With this substitution, Eq. \eqref{eq:55} becomes
\begin{equation}
\frac{d^{2}\mathcal{J}\left(\mathcal{U}\right)}{d\mathcal{\mathcal{U}}^{2}}+\left[-\frac{1}{4}+\frac{\delta}{4M\omega\mathcal{\mathcal{U}}}+\frac{\frac{1}{4}-\left(\frac{\vartheta}{2}\right)^{2}}{\mathcal{\mathcal{U}}^{2}}\right]\mathcal{J}\left(\mathcal{U}\right)=0.\label{eq:57}
\end{equation}
It can be shown that Eq. (\ref{eq:57}) is of the confluent hypergeometric type, whose solution is given in terms of Kummer functions. Thus, using the definition \eqref{eq:32}, the solution of Eq. (\ref{eq:57}) can be written as
\begin{equation}
\psi\left(\boldsymbol{r}\right)=|\mathcal{C}_{3}|\left(\begin{array}{c}
1+\frac{E}{\mathcal{N}}\\
1-\frac{E}{\mathcal{N}}
\end{array}\right)\left(M\omega r^{2}\right)^{\frac{\left|\vartheta\right|}{2}}e^{-\frac{M\omega}{2}r^{2}}e^{-i\left(Et-j\varphi-ik_{z}z\right)}\, _{1}F_{1}\left(\frac{|\vartheta|}{2}-\frac{\delta}{4m\omega}+\frac{1}{2},|\vartheta|+1,M\omega r^{2}\right),\label{eq:60}
\end{equation}
where the parameters $\vartheta$ and $\delta$ are defined in Eq. \eqref{eq:54}. The energies of the KGO are found by the relation 
\begin{equation}
\frac{|\vartheta|}{2}-\frac{\delta}{4M\omega}+\frac{1}{2}=-n.\label{cnd}
\end{equation}
\begin{figure}[!t]
\centering
\includegraphics[scale=0.35]{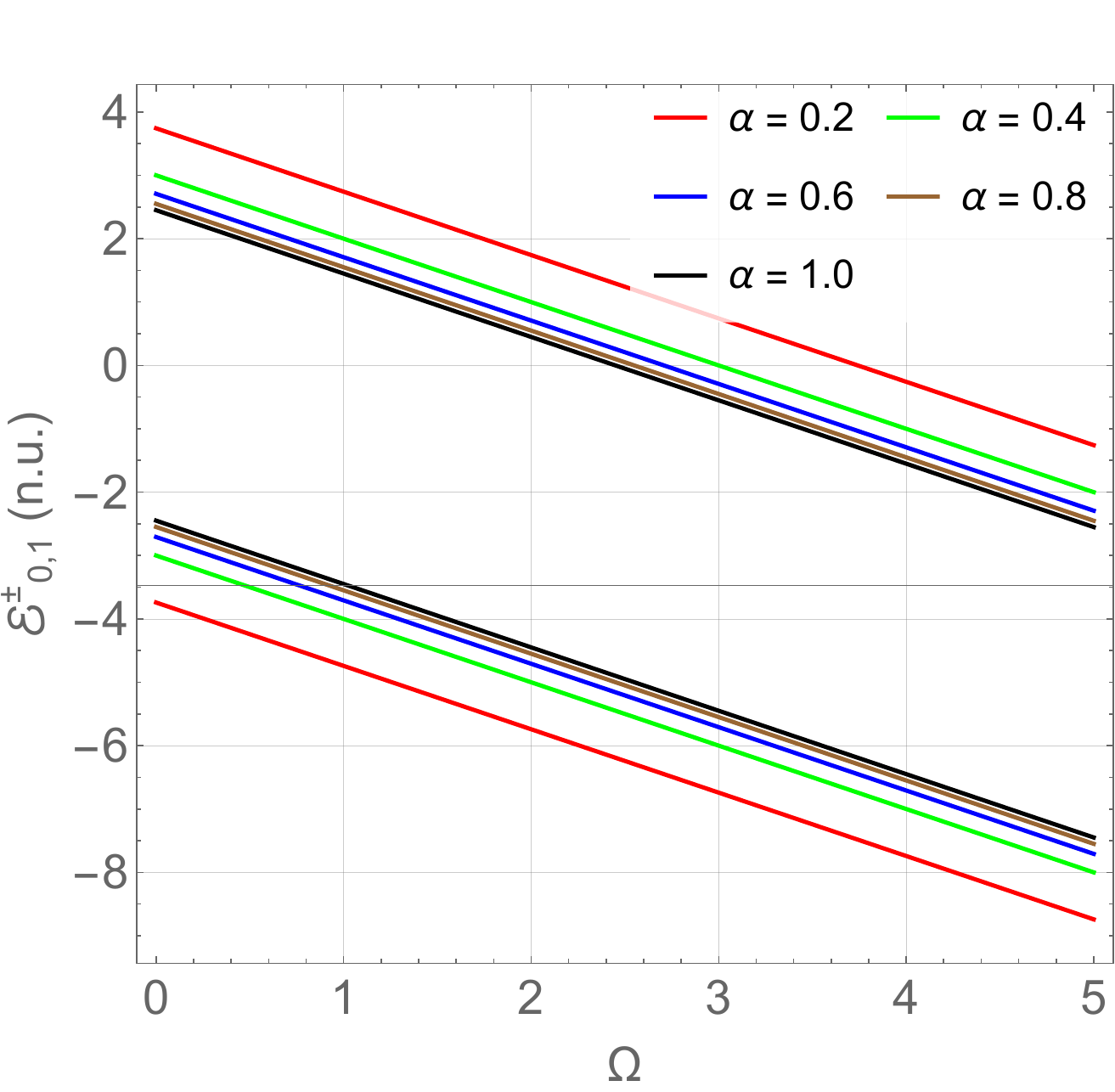}
\caption{Sketch of the energies (Eq. (\ref{energy_KG})) as a function of $\Omega$ for different values of $n$ and $\alpha$. We use $M=1$, $k_{z}=1$, and $\omega=1$.}
\label{Fig_KG_1}	
\end{figure}\begin{figure}[!t]
\centering
\includegraphics[scale=0.35]{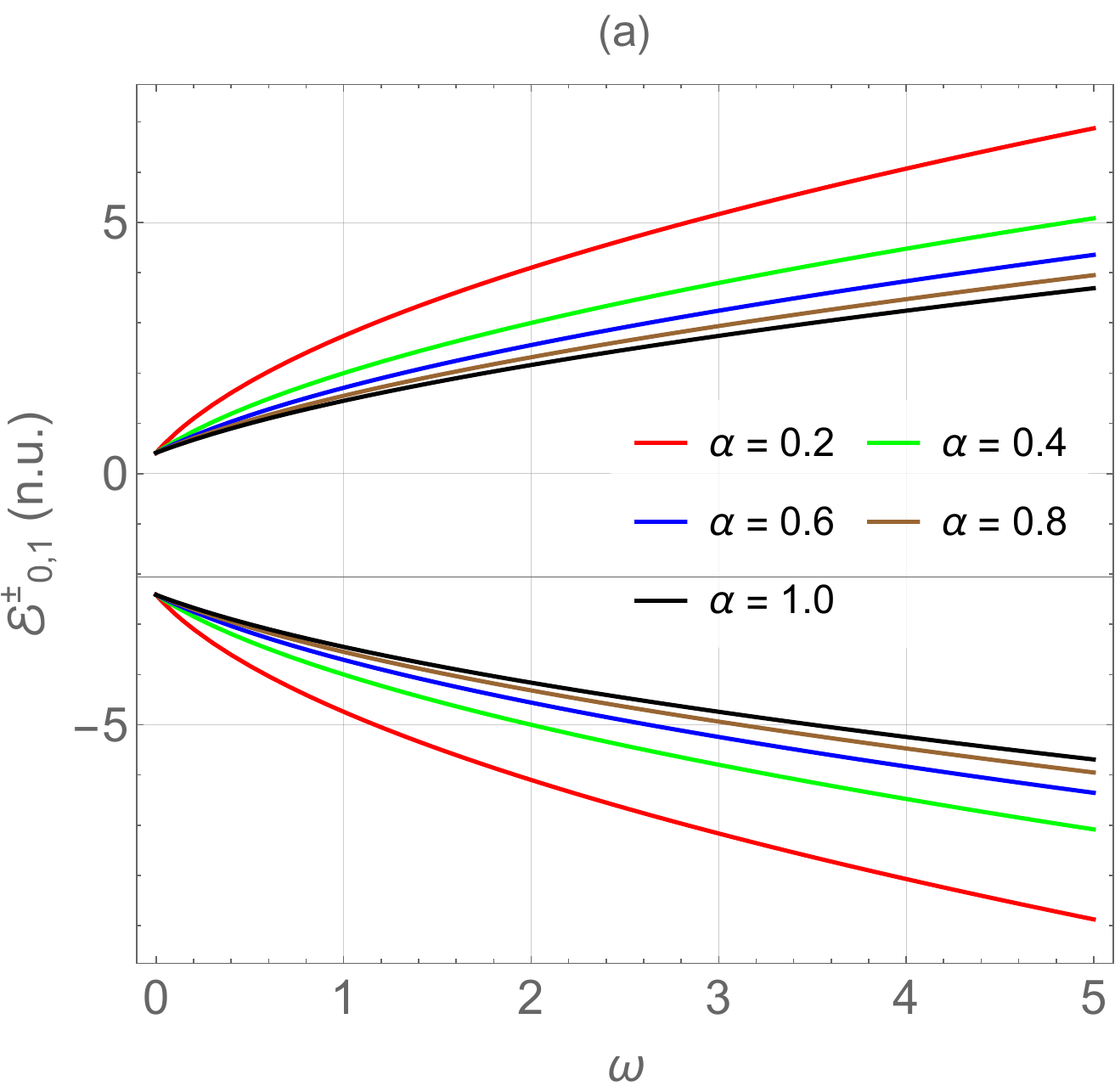}
\includegraphics[scale=0.35]{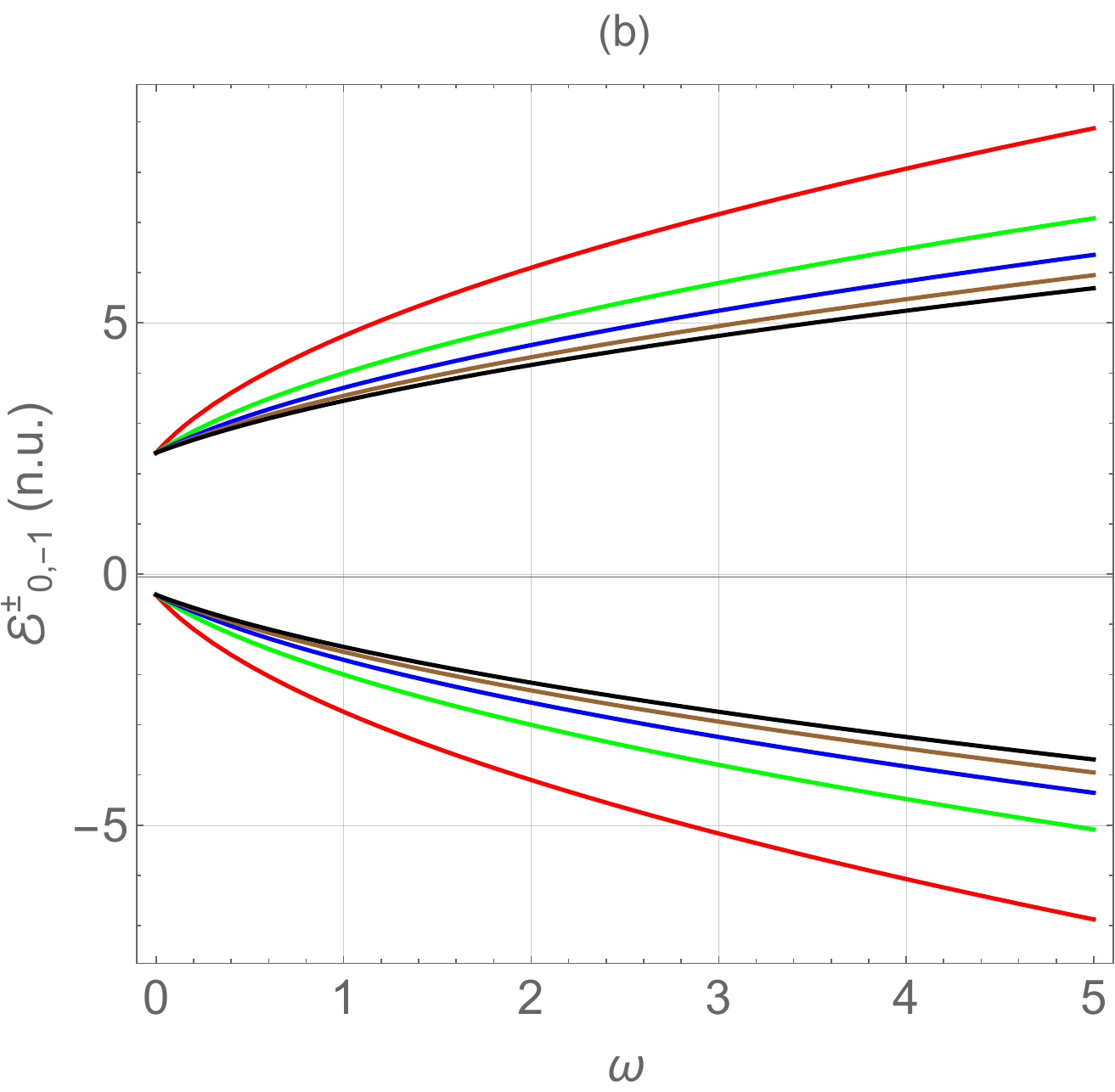}
\caption{Sketch of the energies (Eq. (\ref{energy_KG})) as a function of $\omega$ for different values of $\alpha$. In (a), we plot for $n=0, j=1$ and {b} $n=0, j=-1$. We use $M=1$, $k_{z}=1$, and $\Omega=1$.}
\label{FIG_KG_BC}	
\end{figure}
\begin{figure}[!t]
\centering
\includegraphics[scale=0.35]{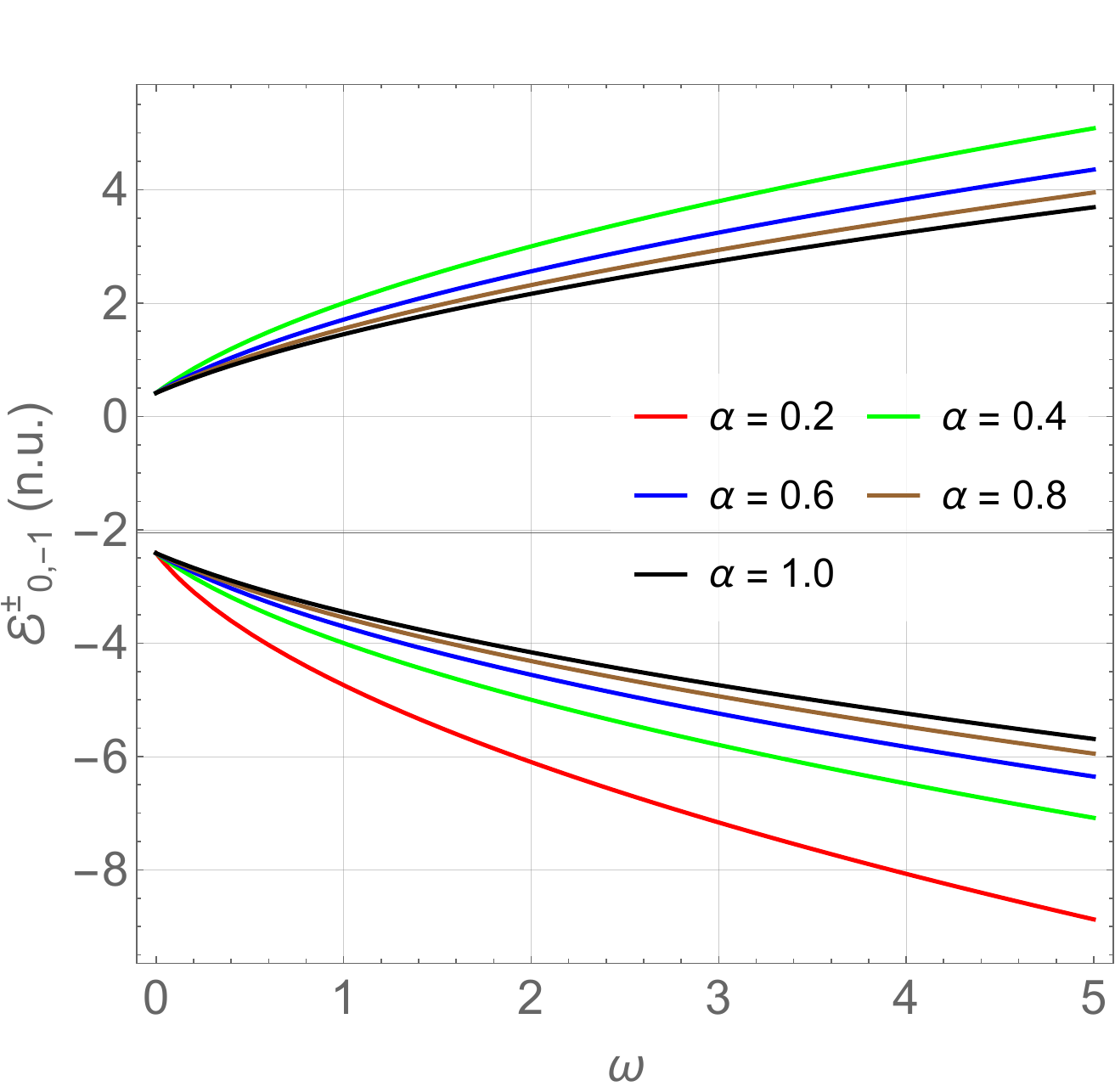}
\caption{Sketch of the energies (Eq. (\ref{energy_KG})) as a function of $\omega$ for $n=0, j=-1, \Omega=-1$ and different values of $\alpha$. We use $M=1$, $k_{z}=1$.}
\label{Fig_KG_D}	
\end{figure}
Substituting back the parameters $\vartheta$ and $\delta$ in Eq.
(\ref{cnd}) and solving the resulting equation for $E$, we obtain the energy levels
\begin{equation}
E^{\pm }(n)=-\Omega j\pm \sqrt{2M\omega \left( 2n+\frac{\left\vert
j\right\vert }{\alpha }+1\right)+M^{2}+k_{z}^{2}}\,, \label{energy_KG}
\end{equation}
and the corresponding wave functions are
\begin{equation}
\psi\left(\boldsymbol{r}\right)=|\mathcal{C}_{3}|\left(\begin{array}{c}
1+\frac{E}{\mathcal{N}}\\
1-\frac{E}{\mathcal{N}}
\end{array}\right)\left(M\omega r^{2}\right)^{\frac{\left|\vartheta\right|}{2}}e^{-\frac{M\omega}{2}r^{2}}e^{-i\left(Et-j\varphi-ik_{z}z\right)}\, _{1}F_{1}\left(-n,|\vartheta|+1,M\omega r^{2}\right).\label{eifKG}
\end{equation}
It is immediate to see that the plot of (\ref{energy_KG}) as a function of $\Omega$ is linear. For the particular state $\mathcal{E}_{0,1}^{(\pm )}$ with $\omega=1$, we find that $\left\vert \mathcal{E}_{0,1}^{(\pm )}\right\vert $ increases when $\alpha$ decreases. On the other hand, when we analyze the profile of (\ref{energy_KG}) as a function of the oscillator frequency $\omega$, we find that $\left\vert \mathcal{E}_{0,1}^{(\pm )}\right\vert $ e $\left\vert \mathcal{E}_{0,-1}^{(\pm )}\right\vert$ increase with $\alpha $ decreasing (Fig. \ref{FIG_KG_BC}). Furthermore, $\left\vert \mathcal{E}_{0,1}^{(+)}\right\vert <\left\vert \mathcal{E}_{0,-1}^{(+)}\right\vert $ while $\left\vert \mathcal{E}_{0,1}^{(-)}\right\vert $ $>$ $\left\vert \mathcal{E}_{0,-1}^{(+)}\right\vert $ (see Figs. \ref{FIG_KG_BC}(a)-(b) for comparison). When we assume negative values of $\Omega $ in Fig. \ref{FIG_KG_BC}(b), for example, $\Omega =-1$, the oscillator energies for $\alpha =0.2$ are
or all negative (see the red solid line in Fig. \ref{Fig_KG_D}). The energies of the FV oscillator are affected by the curvature, specifically by the term $|j|/\alpha$. 

\section{Conclusion}\label{sec5}

We have investigated the relativistic quantum motion of a spinless particle by solving the KG equation in the FV representation. We have considered the particle under spin effects in the space-time cosmic string. We briefly reviewed the VF formalism and subsequently used the results to find the KG equation in the space-time of the spinning cosmic string. To illustrate the importance of the model studied, we have performed three applications. First, we study the KG equation for the particular case where the particle is confined to a hard-wall potential, next we study the motion of the particle in the presence of a Coulomb-type potential, and we finish by studying the KG equation with the inclusion of the KGO.
Wave functions and energies for the three models were determined. We have shown through graphical sketches of the energies that for certain values of rotation and curvature (through the $\alpha$ parameter), the particle's energy shows limited profiles, especially with the presence of rotation. We have argued that curvature strongly impacts the magnitude of the energy levels. With these results, we would like to point out that other physical properties can be studied, such as optical and thermodynamic properties and possible connections to other physical systems in condensed matter.


\begin{thebibliography}{99}
\bibitem[1]{key-1}A. Einstein, Ann. Phys. 49, 769 (1916).

\bibitem[2]{key-2}B. P. Abbott et al, Phys Rev Lett 116, 061102 (2016).

\bibitem[3]{key-3}K. Akiyama et al., Astrophys J Lett 875, L1 (2019).

\bibitem{key-4}R. P. Feynman and A. R. Hibbs, Quantum mechanics and
path integrals,Courier Corporation(1965).

\bibitem{key-5}M. D. Schwartz, Quantum field theory and the standard
model, Cambridge university press(2013).

\bibitem{key-6}A. Ashtekar and J. J. Stachel, Conceptual problems
of quantum gravity, Birkhäuser(1991).

\bibitem{key-7}L. Smolin, The trouble with physics: The rise of
string theory, the fall of a science, and what comes next, (2007).

\bibitem{key-8}N. D. Birrell and P. Davies, Quantum fields in curved
space, Cambridge Monographs on Mathematical Physics (1980).

\bibitem{key-9}L. Parker and D. J. Toms, Quantum field theory in
curved spacetime: Quantized fields and gravity, Cambridge university
press. (2009).

\bibitem{key-10}S. W. Hawking, Comm. Math. Phys. 43, 199 (1975).

\bibitem{key-11}W. G. Unruh and R. M. Wald, Phys. Rev. D 25, 942
(1982).

\bibitem{key-12}G. L. Sewell, Ann. Physics 141, 201 (1982).

\bibitem{key-13}T. W. B. Kibble, J. Phys. A 9, 1387 (1976).

\bibitem{key-14}Y. B. Zel’dovich, Mon. Not. R. Astron Soc. 192, 663
(1980).

\bibitem{key-15}A. Vilenkin, Phys. Rep. 121, 263 (1985).

\bibitem{key-16}T. W. B. Kibble, Phys. Rep. 67, 183 (1980).

\bibitem{key-17}A. Vilenkin, Phys. Lett. B 133, 177 (1983).

\bibitem{key-18}A. Vilenkin and E. P. S. Shellard, Cosmic strings
and other topological defects, Cambridge University Press(1985).

\bibitem{key-19}M. Moshinsky and Y. F. Smirnov, The harmonic oscillator in modern physics, Editions Harwood Academic Publishers, Amsterdam (1996).

\bibitem{key-20}D. Itô, K. Mori, and E. W. Carriere, Il Nuovo Cimento A 51, 1119 (1967).

\bibitem{key-21}M. Moshinsky, and A. Szczepaniak, J. Phys. A, Mathematical and General \textbf{22}, L817 (1989).

\bibitem{Silva1}M\'{a}rcio M. Cunha, Henrique S. Dias, and Edilberto O. Silva. Phys. Rev. D \textbf{102}, 105020 (2020).

\bibitem{Silva2}Daniel F. Lima, Fabiano M. Andrade, Luis B. Castro, Cleverson Filgueiras, and Edilberto O. Silva. Eur. Phys. J. C  \textbf{79}, 596 (2019).

\bibitem{Silva3}Fabiano M. Andrade and Edilberto O. Silva. EPL, \textbf{108}, 30003 (2014).

\bibitem{Silva4}Fabiano M. Andrade and Edilberto O. Silva. Physics Letters B \textbf{738}, 44–47 (2014). 

\bibitem{Silva5}F. M. Andrade, E. O. Silva, M. M. Ferreira Jr., and E. C. Rodrigues. Physics Letters B \textbf{731}, 327–330 (2014).

\bibitem{boumali1}A. Boumali. EJTP \textbf{32}, 121-130 (2015).

\bibitem{boumali2}A. Boumali and L. chetouani. Phys. Lett. A \textbf{346}, 261-268 (2005).

\bibitem{key-22}S. A. Bruce and P. C. Minning, Il Nuovo Cimento A
106, 711 (1993).

\bibitem{key-23}V. V. Dvoeglazov, Il Nuovo Cimento A 107, 1785 (1994).

\bibitem{key-24}J. Carvalho, A. M. de M. Carvalho, E. Cavalcante,
and C. Furtado, Eur. Phys. J. C \textbf{76}, 1 (2016).

\bibitem{key-25}L. C. dos Santos and C. de Camargo Barros, Eur. Phys. J. C \textbf{78}, 1 (2017).

\bibitem{key-26}R. L. L. Vitória and K. Bakke, Eur. Phys. J. C \textbf{78}, 1 (2018).

\bibitem{key-27}R. R. Cuzinatto, M. de Montigny, and P. Pompeia,
Class. Quan. Grav 39,075006 (2022).

\bibitem{key-28}F. Ahmed, Europhys. Lett. \textbf{131}, 30002 (2020).

\bibitem{key-29}K. M. Case, Phys. Rev. \textbf{95}, 1323 (1954).

\bibitem{key-30}L. L. Foldy, Phys. Rev. \textbf{102}, 568 (1956).

\bibitem{key-31}L. L. Foldy and S. A. Wouthuysen, Phys. Rev. \textbf{78}, 29 (1950).

\bibitem{key-32}H. Feshbach and F. M. H. Villars, Rev. Modern Phys.
\textbf{30}, 24 (1958).

\bibitem{key-33}B. A. Robson and D. S. Staudte, J. Phys. A: Math.Gen 29, 157 (1996).

\bibitem{boumali3}A. Boumali and H. Aounallah, Advances in High Energy Physics \textbf{2018} (2018).

\bibitem{boumali4}H. Aounallah, A. Boumali. Phys. Part. Nuclei Lett. \textbf{16}, 195-205 (2019).

\bibitem{boumali5}A. Boumali and H. Aounallah, Rev. Mex. Fis. \textbf{66}, 192-208 (2020).

\bibitem{key-34}D. S. Staudte, J. Phys. A \textbf{29}, 169 (1996).

\bibitem{key-35}M. Merad, L. Chetouani, and A. Bounames, Phys. Lett.
A \textbf{267}, 225 (2000).

\bibitem{key-36}A. Bounames and L. Chetouani, Phys. Lett. A \textbf{279}, 139 (2001).

\bibitem{key-37}S. Haouat and L. Chetouani, Eur. Phys. J. C \textbf{41}, 297 (2005).

\bibitem{key-38}N. Brown, Z. Papp, and R. M. Woodhouse, Few-Body
Systems \textbf{57}, 103 (2015).

\bibitem{key-39}B. Motamedi, T. Shannon, and Z. Papp, Few-Body Systems, \textbf{}, 1-7 (2019).

\bibitem{key-40}O. Klein, Z. Phys \textbf{37}, 895 (1926).

\bibitem{key-41}W. Gordon, Z. Phys \textbf{40}, 117 (1926).

\bibitem{key-42}W. Greiner, Relativistic quantum mechanics. wave
equations, 2000.

\bibitem{key-43}F. L. Gross, Relativistic quantum mechanics and field theory, John Wiley and Sons (1993).

\bibitem{key-44}A. J. Silenko, Phys. Rev. A \textbf{77}, 012116 (2008).

\bibitem{key-45}B. Mirza and M. Mohadesi, Commun. Theor. Phys \textbf{42}, 664 (2004).

\bibitem{key-46}I. Gott, J. R., Astrophys J \textbf{288}, 422 (1985).

\bibitem{key-47}Hiscock, Phys. Rev. D \textbf{31}, 3288 (1985).

\bibitem{key-48}A. Bouzenada, A. Boumali, Ann. Physics \textbf{452}, 169302 (2023).

\bibitem{key-49}A. Bouzenada, A. Boumali and M. Al-Raeei, 2023. preprint arXiv:2302.13805.

\bibitem{key-50}J. R. Gott and M. Alpert, Gen. Rel. Grav. \textbf{16}, 243 (1984).

\bibitem{key-51}Gal’tsov and Letelier, Phys. Rev. D \textbf{47}, 4273 (1993).

\bibitem[52]{key-52}A. Boumali and N. Messai, Can. J. Phys. \textbf{92}, 1460 (2014).

\bibitem[53]{key-m53}A. Vilenkin, Phys. Rep. \textbf{121}, 263 (1985).

\bibitem[54]{key-m54}K. Bakke, Eur. Phys. J. Plus \textbf{127}, 82 (2012).

\bibitem[55]{key-m55}K. Bakke, C. Furtado, Eur. Phys. J. C \textbf{69}, 531 (2010).

\bibitem[56]{key-m56}K. Bakke, C. Furtado, Phys. Rev. D \textbf{80}, 024033 (2009).

\bibitem[57]{key-m57}K. Bakke, C. Furtado, Phys. Rev. D \textbf{82}, 084025 (2010).

\bibitem[58]{key-m58}T. W. B. Kibble, J. Phys. A \textbf{9}, 1387 (1976).

\bibitem[59]{key-m59}P. O. Mazur, Phys. Rev. Lett. \textbf{57}, 929 (1986).


\bibitem{key-53}A. J. Silenko, Theoret. Math. Phys. \textbf{156}, 1308 (2008).

\bibitem{key-54}A. J. Silenko, Phys. Rev. D \textbf{88}, 045004 (2013).

\bibitem{key-55}A. Mostafazadeh, J. Phys. A \textbf{31}, 7829 (1998).

\bibitem{key-56}Z. Soroush, H. Hassanabadi, and Marc de Montigny. General Relativity and Gravitation 52,1-20(2020).
\end{thebibliography}
\end{document}